\documentclass[12pt]{article}
\pdfoutput=1
\usepackage{mathtools,amssymb,mathrsfs,microtype,tikz,ytableau}
\usepackage{tikz-cd}
\usepackage{tabularx,ragged2e}
\usepackage[utf8]{inputenc}
\usepackage{float}
\usepackage{xcolor}
\usepackage{empheq}
\usepackage{amsmath}
\usepackage{slashed}
\usepackage{caption}
\usepackage{dsfont}
\usepackage{verbatim}
\usepackage{tcolorbox}
\usepackage{subfig}
\usepackage{amsfonts}
\usepackage{graphicx}
\usepackage{physics}
\usepackage{slashed}
\usetikzlibrary{decorations.pathmorphing}
\usepackage[colorlinks=true]{hyperref}
\DeclarePairedDelimiter{\ceil}{\lceil}{\rceil}
\newcolumntype{C}{>{\Centering\arraybackslash}X}

\usepackage[backend=bibtex,style=numeric-comp,sorting=none,maxbibnames=99, minbibnames=99]{biblatex}	
\ytableausetup{centertableaux,boxsize=.5em}
\makeatletter\@addtoreset{equation}{section}\makeatother

\DeclareMathOperator{\sign}{sign}

\renewcommand{\title}[1]{\vbox{\center\LARGE{#1}}\vspace{5mm}}
\renewcommand{\author}[1]{\vbox{\center\large#1}\vspace{5mm}}
\newcommand{\address}[1]{\vbox{\center\em#1}}

\newcommand{\beq}{\begin{equation}\begin{aligned}}
\newcommand{\eeq}{\end{aligned}\end{equation}}

\newcommand{\ti}{\frac{2i}{3}}
\newcommand{\tp}{\frac{1}{2\pi}}

\newcommand{\fp}{\frac{1}{4\pi}}

\newcommand{\ov}{\over}

\begin{document}
 
\begin{titlepage}
\begin{center}
\hfill {\tt }\\
\vspace{1mm}

\title{
 {\LARGE Phases of Two Adjoints QCD$_{3}$ And a Duality Chain}}
\vspace{4mm}

Changha Choi,${}^{ab}$\footnote{\href{mailto: changha.choi@stonybrook.edu}
{\tt changhachoi@gmail.com}}
\vskip 3mm
\address{
${}^a$Physics and Astronomy Department,\\ Stony Brook University, Stony Brook, NY 11794, USA
}

\address{
${}^b$Simons Center for Geometry and Physics,\\ Stony Brook, NY 11794, USA
}

\end{center}

\vspace{5mm}
\abstract{
\noindent
We analyze the 2+1 dimensional gauge theory with two fermions in the real adjoint representation with non-zero Chern-Simons level.  We propose a new fermion-fermion dualities between strongly-coupled theories and determine the quantum phase using the structure of a \emph{`Duality Chain'}. We argue that when Chern-Simons level is sufficiently small, the theory in general develops a strongly coupled quantum phase described by an emergent topological field theory. For special cases, our proposal predicts an interesting dynamical scenario with spontaneous breaking of partial 1-form or 0-form global symmetry. It turns out that $SL(2,\mathbb Z)$ transformation and the generalized level/rank duality are crucial for the unitary group case. We further unveil the dynamics of the 2+1 dimensional gauge theory with any pair of adjoint/rank-two fermions or two bifundamental fermions using similar \emph{`Duality Chain'}. 
}
\vfill

\end{titlepage}

{\hypersetup{linkcolor=black}
\tableofcontents
\thispagestyle{empty}
}

\section{Introduction}
\label{sec:intro}
\setcounter{footnote}{0}

In this paper, we analyze the low-energy dynamics of Two Adjoints QCD$_3$ in 2+1 dimensional quantum field theory. The theory we consider consists of Yang-Mills term with gauge group $G=SU(N),SO(N),Sp(N)$ with Chern-Simons level $k$, together with two real adjoint fermions or equivalently single Dirac adjoint fermion.\footnote{We denote `adjoint fermion' as a Majorana fermion in the adjoint representation of the gauge group. In the real basis $\gamma^0=i\sigma^2,~\gamma^1=\sigma^1,\gamma^2=\sigma^3$, Majorana spinor is equivalent to real spinor.} We construct a phase diagram as a function of the bare mass parameter $m$.

When the Chern-Simons level $k$ is large enough, semiclassical analysis predicts phase diagram with two asymptotic regions distinguished by a single phase transition\cite{Avdeev:1991za,Avdeev:1992jt}. Simple dimensional analysis indicates that IR dynamics become strongly coupled for a sufficiently small Chern-Simons level $k$, and many recent developments reviewed below indicate the existence of a non-perturbative quantum phase in the infrared when the bare quark mass is sufficiently small. As a consequence, there are two different phase transitions upon varying the mass of fermions from negative to positive infinity. So far, the standard route to obtain a phase diagram consistent with various anomalies comes from finding two mutually non-local dual descriptions with common global symmetry\cite{Komargodski:2017keh,Gomis:2017ixy,Cordova:2017vab,Cordova:2017kue,Choi:2018tuh,Cordova:2018qvg,Aitken:2019mtq}. Particularly, all these dual descriptions are in the semiclassical regime, in the sense that intermediate quantum phase could be accessed by weakly coupled analysis of dual descriptions.\footnote{ \label{fn:weakstrong}From now on, we frequently use the terms `weakly-coupled' and `strongly-coupled'. Although interaction at the critical point is not parametrically small in the low-energy regime except for the few limits of the parameter space such as large k, we use the term `weakly-coupled' when the phase diagram of the theory develeops two semiclassical phase with large asymptotic mass separated by single transition point between them. The term `strongly-coupled' is for the case other than `weakly-coupled'.}

Up to now nonperturbative quantum phase was investigated for the case of fermions in the fundamental, single real adjoint and sym/asym representations. The next natural generalization one could ask is to find a phase diagram when the matter content is two adjoint fermions since the existence of quantum phase was indicated in the case of $SU(2)$\cite{Choi:2018tuh} together with the various hints from the 3+1 dimensional physics with single Dirac adjoint fermion\cite{Anber:2018tcj,Bi:2018xvr,Cordova:2018acb,Wan:2018djl}. But we were unable to replicate the previously successful apporach. Instead, we found a new structure which we dub \emph{`Duality Chain'}. This allows to compute the quantum phase, but not with a weakly-coupled dual description.

To see more clearly the contribution of this paper, it is worth reviewing at least briefly the recent progress in the non-supersymmetric dualities in 2+1 dimensions so far relevant to our paper. Originally motivated from AdS/CFT, Chern-Simons matter theories and its bosonization duality was suggested and supported by many exact computations in the ’t Hooft limit or flows from the $\mathcal N=2$ supersymmetric dualities (large $k,N$ with fixed $\lambda=N/k$)\cite{Giombi:2011kc,Aharony:2011jz,Aharony:2012nh,GurAri:2012is,Aharony:2012ns,Jain:2013py,Jain:2013gza,Inbasekar:2015tsa,Gur-Ari:2015pca,Aharony:2018pjn}. On the other hand, motivated from the recent studies of quantum hall effect and topological insulator, generalization of bosonic particle-vortex\cite{Peskin:1977kp,Dasgupta:1981zz} duality and the webs connecting them were discovered for the case of abelian gauge group \cite{Son:2015xqa,Metlitski:2015eka,Xu:2015lxa,Metlitski:2016dht,Mross:2015idy,Seiberg:2016gmd,Karch:2016sxi,Murugan:2016zal,Karch:2016aux}. Generalizations of bosonization duality\cite{Radicevic:2015yla,Aharony:2015mjs,Hsin:2016blu, Aharony:2016jvv} to the non-abelian gauge theory with finite $N$ and $k$ using the exact level/rank dualities were conjectured with pieces of evidence supporting them, e.g. \cite{Radicevic:2016wqn,Jensen:2017xbs,Benini:2017dus,Giombi:2017txg,Aitken:2018joi}. See also \cite{Chester:2017vdh,Benini:2017aed,Jensen:2017bjo,Chen:2018vmz,Benvenuti:2018cwd,Benvenuti:2019ujm,Sharon:2018apk,Bashmakov:2018ghn,DiPietro:2019hqe,Kan:2019rsz} for the recent developments in the non-supersymmetric gauge theories in 2+1 dimensions. But there were restrictions on the parameter space depending on the number of flavors and Chern-simons level which called `flavor-bound' in \cite{Aitken:2018cvh,Aitken:2019mtq}. Extension of the bosonization duality to the all possible parameter space led to conjecture the existence of the non-perturbative quantum phase near the massless regime similar to the $\mathcal N=1$ supersymmetric theories \cite{Bashmakov:2018wts,Choi:2018ohn,Aharony:2019mbc}. Now the bosonization is carried out at the two distinguished critical points with two different mutually non-local dual descriptions \cite{Komargodski:2017keh}. This shape of the phase diagram has been successfully applied to the case of single adjoint/rank-two matter and some other cases such as single bifundamental fermions \cite{Gomis:2017ixy, ,Cordova:2017vab, Choi:2018tuh, Cordova:2018qvg, Aitken:2019mtq, Aitken:2019shs} with various supporting checks \cite{Hong:2010sb,Argurio:2018uup,Armoni:2019lgb,Karthik:2018nzf,Armoni:2017jkl,Akhond:2019ued,Kanazawa:2019oxu}. (See \cite{Turner:2019xhl} for the recent lecture note covering many aspects discussed above.) It is important to note that all these mutually non-local dual descriptions were weakly coupled in the sense that there are only two semiclassically-accessible IR phase distinguished by a single critical point.

In this respect, we employ the possibility of strongly coupled dual description to construct a consistent phase diagram of Two Adjoints QCD$_3$ and its cousins. Our dual descriptions are generically strongly coupled theories in the sense of footnote \ref{fn:weakstrong} thus the quantum phase of the original theory is not directly accessible. Now, the concept of \emph{`Duality Chain'} is introduced as an algorithm for the determination of the quantum phase. Since original and strongly coupled dual theory share the same quantum phase, there exists a dual of a dual description which flows to the same quantum phase. If this is weakly-coupled, then we land on our feet and the quantum phase can be determined semi-classically. If it is still strongly-coupled, then we run the same procedure recursively in a sense that we keep investigating the dual of the strongly coupled theories which shares the common intermediate quantum phase. Remarkable observation for Two Adjoints QCD$_3$ is that as we go over this chain of dualities, weakly coupled description of the initial quantum phase can be obtained within a finite number of steps for the non-zero Chern-Simons level $k$. We specify the duality chain and the phases of two adjoints fermions under the $SU/SO/Sp$ gauge group which are consistent under various non-trivial test. We emphasize that duality chain is not applicable anymore for the case of vanishing Chern-Simons level $k=0$ and we comment the problem in the section \ref{sec:comments}.

We also find that \emph{`Duality Chain'} is successfully applicable for the following two types of theories. One is the dynamics of QCD$_3$ with arbitrary pair of fermions in the adjoint/rank-two representations which is the full generalization of the QCD$_3$ with single flavor fermion in adjoint/rank-two representation\cite{Gomis:2017ixy,Choi:2018tuh}. The other is the dynamics of the 2-node quiver gauge theory with two bifundamental fermions which is the generalization of the QCD$_3$ with the vector fermions\cite{Komargodski:2017keh} by gauging flavor symmetry\cite{Aitken:2019shs}. It is important to note that the latter model with two bifundamental fermions passes more stringent consistency checks due to its distinguished characteristics and we propose possible 2+1d orbifold equivalence \cite{Jensen:2017dso, Aitken:2019shs} between two theories at the end of this paper.

\vspace{8pt}

Let us now summarize the main proposals regarding Two Adjoints QCD$_3$:

\begin{enumerate}

\item These theories have a critical value of the level $k_\mathrm{crit}$ below which a new intermediate quantum phase appears between the semiclassically accessible asymptotic large-positive and large-negative mass phases. The critical value is given by the dual coxeter number of gauge group G.
\begin{equation}
k_\mathrm{crit}=h
\end{equation}

\item For $k\geq k_\mathrm{crit}$\footnote{From now on, we set $k\geq 0$ without loss of generality.}, the phase diagram has just two phases: the asymptotic large-positive mass and large-negative mass semiclassical phases, separated by a phase transition. For very large $k$ the phase transition is controlled by a weakly-coupled CFT. The asymptotic large mass phases are the TQFTs
\begin{equation}
G_{k\pm h}
\label{intermed}
\end{equation}
where the upper/lower sign is for the large positive/negative mass asymptotic phase. These phases are present in these theories for any $k$.

\item For $0<k<k_\mathrm{crit}$, the theories undergo two phase transitions as a function of the mass of the fermions. Two distinct phase transitions connect the intermediate quantum phase with the asymptotic large-positive mass phase and with 
the asymptotic large-negative mass phase. Let's denote the location of two critical points as $ m^{\pm crit}$ ($m^{+crit}>m^{-crit}$). Then we propose new (fermion-fermion) dualities in $2+1$ dimensions at each critical point. This duality is conceptually different compared with previously conjectured non-supersymmetric dualities in $2+1$ dimensions because it's generically between the strongly-coupled theories which have non-perturbative quantum phases.

 \smallskip
Dualities for $SU(N)_k+~\text{2}~ \psi^{adj}$ for $0<k<N$:

\begin{equation}
\begin{aligned}
&SU(N)_k+~\text{2}~ \psi^{adj},~ m_\psi=m^{+crit}_\psi &\longleftrightarrow~~ &U\!\left(N+k\right)_{k,-N}+~\text{2}~ \tilde \psi^{adj},~ m_{\tilde \psi}=m^{-crit}_{\tilde \psi} \\[+5pt]
&SU(N)_k+~\text{2}~ \psi^{adj},~ m_{\psi}=m^{-crit}_{\psi} &\longleftrightarrow~~ &U\!\left(N-k\right)_{k,N}+~\text{2}~\hat \psi^{adj},~ m_{\hat \psi}=m^{+crit}_{\hat \psi}\,.
\end{aligned}
\end{equation}

\medskip

    Dualities for $SO(N)_k+~\text{2}~ \psi^{adj}$ for $0<k<N-2$:
\begin{equation}
\begin{aligned}
&SO(N)_k+~\text{2}~ \psi^{adj},~ m_\psi=m^{+crit}_\psi &\longleftrightarrow ~~&SO\!\left(N+k-2\right)_{k}+~2~ \tilde \psi^{sym},~ m_{\tilde \psi}=m^{-crit}_{\tilde \psi} \\[+5pt]
&SO(N)_k+~\text{2}~ \psi^{adj},~ m_{\psi}=m^{-crit}_\psi &\longleftrightarrow~~& SO\!\left(N-k-2\right)_{k}+~2~ \hat \psi^{sym},~ m_{\hat \psi}=m^{+crit}_{\hat \psi}\,.
\end{aligned}
\end{equation}

\medskip

Dualities for $Sp(N)_k+~\text{2}~ \psi^{adj}$ for $0<k<N+1$:
\begin{equation}
\begin{aligned}
&Sp(N)_k+~\text{2}~ \psi^{adj},~ m_\psi=m^{+crit}_\psi &\longleftrightarrow~~& Sp\!\left(N+k+1\right)_{k}+~2~ \tilde \psi^{asym},~ m_{\tilde \psi}=m^{-crit}_{\tilde \psi} \\[+5pt]
&Sp(N)_k+~\text{2}~ \psi^{adj},~ m_{\psi}=m^{-crit}_\psi &\longleftrightarrow~~& Sp\!\left(N-k+1\right)_{k}+~2~ \hat \psi^{asym},~ m_{\hat \psi}=m^{+crit}_{\hat \psi}\,.
\end{aligned}
\end{equation}

We note that for the $SO/Sp$ gauge theory, fermions in the dual gauge theory transforms in the other rank-two representation compared to the fermions in the original gauge theory which is similar to the single flavor case \cite{Gomis:2017ixy}. To complete the phase diagram of $SO/Sp$ gauge theory, the following additional dualities are crucial :

\smallskip

Dualities for $SO(N)_k+~\text{2}~ \psi^{sym}$ for $0<k<N-2$:
\begin{equation}
\begin{aligned}
&SO(N)_k+~\text{2}~ \psi^{sym},~ m_\psi=m^{+crit}_\psi &\longleftrightarrow~~& SO\!\left(N+k+2\right)_{k}+~2~ \tilde \psi^{adj},~ m_{\tilde \psi}=m^{-crit}_{\tilde \psi} \\[+5pt]
&SO(N)_k+~\text{2}~ \psi^{sym},~ m_{\psi}=m^{-crit}_\psi &\longleftrightarrow~~& SO\!\left(N-k+2\right)_{k}+~2~ \hat \psi^{adj},~ m_{\hat \psi}=m^{+crit}_{\hat \psi}\,.
\end{aligned}
\end{equation}

\medskip

Dualities for $Sp(N)_k+~\text{2}~ \psi^{asym}$ for $0<k<N-1$:
\begin{equation}
\begin{aligned}
&Sp(N)_k+~\text{2}~ \psi^{asym},~ m_\psi=m^{+crit}_\psi &\longleftrightarrow~~& Sp\!\left(N+k-1\right)_{k}+~2~ \tilde \psi^{adj},~ m_{\tilde  \psi}=m^{-crit}_{\tilde \psi} \\[+5pt]
&Sp(N)_k+~\text{2}~ \psi^{asym},~ m_{\psi}=m^{-crit}_\psi &\longleftrightarrow~~& Sp\!\left(N-k-1\right)_{k}+~2~ \hat \psi^{adj},~ m_{\hat \psi}=m^{+crit}_{\hat \psi}\,.
\end{aligned}
\end{equation}

\item The quantum phase of the strongly coupled theory $\mathcal C_0$ is obtained by using the concept of \emph{`Duality Chain'} as illustrated in figure \ref{fig:dualitychain}. We already described the two mutually non-local descriptions share the same quantum phase above, and we call them $\mathcal C_{-1}$ and $\mathcal C_{+1}$ as in the figure. We define that $\mathcal C_1/\mathcal C_{-1}$ has higher/lower \emph{`rank'} than $\mathcal C_0$. Now, if $\mathcal C_{-1}$ is still strongly coupled, there must be another dual description describing the same quantum phase with smaller rank denoted as $\mathcal C_{-2}$. Using this step recursively, we would obtain the weakly coupled theory describing quantum phase $Q[\mathcal {C}_0]$ within in finite $n^*$ steps. We emphasize that critical point of final weakly coupled description of the quantum phase $\mathcal C_{-n^*}$ is in general not dual to any of the original $\mathcal C_0$'s critical points.

\item One could also go up the \emph{`Duality Chain'} and thus obtain the infinite number of theories $\mathcal C_i,~ i>-n^*$ sharing the same quantum phase as in figure \ref{fig:dualitychain}.

\item Our quantum phase $Q[\mathcal C_0]$ of Two Adjoints QCD$_3$ is generally Chern-Simons TQFT. But in some special cases, it turns out that quantum phases are TQFT with partially spontaneously-broken one-form or zero-form global symmetry. We emphasize that this non-trivial scenario of partial deconfinement in QCD$_3$ is special feature of two adjoints QCD$_3$ which is absent in the single adjoint QCD$_3$ case of \cite{Gomis:2017ixy}.

\item Finally, analysis of the 2 adjoint fermions case naturally lead us to conjecture similar duality chain structure is inherent for QCD$_3$ with any two combination of rank-two/adjoint fermions or the theory of two bifundamental fermions. We enumerate such generalizations in section \ref{sec:more dualities}.

\end{enumerate}

\begin{figure}[!h]
\makebox[\textwidth][c]{
\begin{tikzpicture}
\pgfmathsetmacro{\M}{7.2}
\pgfmathsetmacro{\L}{3.3}
\pgfmathsetmacro{\H}{3.0}


\begin{scope}
\clip (\L,-2*\H-.1) rectangle (\M+.1,.1);
\draw[thick,->,>=stealth,line width=0.5mm] (-\M,0) -- (\M,-0.5*\H) -- (-\M,-\H)  -- (\M,-1.5*\H)-- (-\M,-2*\H) -- (\M,-2.5*\H);
\end{scope}

\begin{scope}
\clip (-\L,-3*\H-.1) rectangle (-\M-.1,\H+.1);
\draw[black, thick,line width=0.5mm]  (\M,0.5*\H) -- (-\M,0) -- (\M,-0.5*\H) -- (-\M,-\H)  -- (\M,-1.5*\H)-- (-\M,-2*\H) --(\M, -2.5*\H );
\end{scope}

\begin{scope}
\clip (-\L,-2.17*\H-.1) rectangle (\L,0.22*\H+.1);
\draw[dashed,thick,line width=0.5mm] (\M,0.5*\H) --  (-\M,0) -- (\M,-0.5*\H) -- (-\M,-\H)  -- (\M,-1.5*\H)-- (-\M,-2*\H) -- (\M,-2.5*\H);
\end{scope}

\begin{scope}
\clip (\L,-4*\H-.1) rectangle (\M+.1,.1);
\draw[thick,->,>=stealth,line width=0.5mm] (-\M,-2.25*\H)--(\M,-2.75*\H)--(-\M, -3.25*\H);
\end{scope}

\begin{scope}
\clip (0.5*\L,-3*\H-.1) rectangle (\L,.1);
\draw[dashed,thick,line width=0.5mm] (-\M,-2.25*\H)--(\M,-2.75*\H);
\end{scope}

\begin{scope}
\clip (0,-3*\H-.1) rectangle (\L,.1);
\draw[dashed,thick,line width=0.5mm] (\M,-2.75*\H)--(-\M, -3.25*\H);
\end{scope}

\fill[red!20!white, fill opacity=0.3]  (-\L, -3.5*\H) rectangle  (\L, 0.5*\H) ;

\fill[blue, fill opacity=0.2]  (-\M-1,+0.25*\H *\L / \M-0.25*\H -0.7) rectangle  (-\L,-0.25*\H *\L / \M+0.25*\H +0.7) ;
\draw[blue,decorate, decoration={snake},line width=0.9mm,opacity=1] (-\L,+0.25*\H *\L / \M-0.25*\H -0.7)--(-\L,-0.25*\H *\L / \M+0.25*\H +0.7);

\fill[blue, fill opacity=0.2]  (-\M-1,-\H+0.25*\H *\L / \M-0.25*\H -0.7) rectangle  (-\L,-\H-0.25*\H *\L / \M+0.25*\H +0.7) ;
\draw[blue,decorate, decoration={snake},line width=1.2mm,opacity=1] (-\L,-\H+0.25*\H *\L / \M-0.25*\H -0.7)--(-\L,-\H-0.25*\H *\L / \M+0.25*\H +0.7);

\fill[blue, fill opacity=0.2]  (-\M-1,-2*\H+0.25*\H *\L / \M-0.25*\H -0.7) rectangle  (-\L,-2*\H-0.25*\H *\L / \M+0.25*\H +0.7) ;
\draw[blue,decorate, decoration={snake},line width=1.2mm,opacity=1] (-\L,-2*\H+0.25*\H *\L / \M-0.25*\H -0.7)--(-\L,-2*\H-0.25*\H *\L / \M+0.25*\H +0.7);

\fill[blue, fill opacity=0.2]   (\L,-0.5*\H+0.25*\H *\L / \M-0.25*\H -0.7) rectangle  (\M+1,-0.5*\H-0.25*\H *\L / \M+0.25*\H +0.7) ;
\draw[blue,decorate, decoration={snake},line width=1.2mm,opacity=1] (\L,-0.5*\H+0.25*\H *\L / \M-0.25*\H -0.7)--(\L,-0.5*\H-0.25*\H *\L / \M+0.25*\H +0.7);

\fill[blue, fill opacity=0.2]   (\L,-1.5*\H+0.25*\H *\L / \M-0.25*\H -0.7) rectangle  (\M+1,-1.5*\H-0.25*\H *\L / \M+0.25*\H +0.7) ;
\draw[blue,decorate, decoration={snake},line width=1.2mm,opacity=1] (\L,-1.5*\H+0.25*\H *\L / \M-0.25*\H -0.7)--(\L,-1.5*\H-0.25*\H *\L / \M+0.25*\H +0.7);

\fill[blue, fill opacity=0.2]   (\L,-2.75*\H+0.25*\H *\L / \M-0.25*\H -0.7) rectangle  (\M+1,-2.75*\H-0.25*\H *\L / \M+0.25*\H +0.7) ;
\draw[blue,decorate, decoration={snake},line width=1.2mm,opacity=1] (\L,-2.75*\H+0.25*\H *\L / \M-0.25*\H -0.7)--(\L,-2.75*\H-0.25*\H *\L / \M+0.25*\H +0.7);

\draw (-\L, 0.5*\H) -- (-\L, -3.5*\H);
\draw (\L, 0.5*\H) -- (\L, -3.5*\H);

\filldraw[white!20!blue] (-\L,0-0.25*\H *\L / \M+0.25*\H) circle (3pt);\draw (-\L,0-0.25*\H *\L / \M+0.25*\H) circle (3pt);
\filldraw[white!20!blue] (-\L,0+0.25*\H *\L / \M-0.25*\H) circle (3pt);\draw (-\L,0+0.25*\H *\L / \M-0.25*\H) circle (3pt);

\filldraw[white!20!blue] (\L,-0.5*\H -0.25*\H *\L / \M+0.25*\H) circle (3pt);\draw (\L,-0.5*\H -0.25*\H *\L / \M+0.25*\H)  circle (3pt);
\filldraw[white!20!blue] (\L,-0.5*\H +0.25*\H *\L / \M-0.25*\H) circle (3pt);\draw (\L,-0.5*\H+0.25*\H *\L / \M-0.25*\H)  circle (3pt);

\filldraw[white!20!blue] (-\L,-\H -0.25*\H *\L / \M+0.25*\H) circle (3pt);\draw(-\L,-\H -0.25*\H *\L / \M+0.25*\H) circle (3pt);
\filldraw[white!20!blue] (-\L,-\H +0.25*\H *\L / \M-0.25*\H) circle (3pt);\draw(-\L,-\H +0.25*\H *\L / \M-0.25*\H) circle (3pt);

\filldraw[white!20!blue] (\L,-1.5*\H -0.25*\H *\L / \M+0.25*\H) circle (3pt);\draw (\L,-1.5*\H -0.25*\H *\L / \M+0.25*\H)  circle (3pt);
\filldraw[white!20!blue] (\L,-1.5*\H +0.25*\H *\L / \M-0.25*\H) circle (3pt);\draw (\L,-1.5*\H+0.25*\H *\L / \M-0.25*\H)  circle (3pt);

\filldraw[white!20!blue] (-\L,-2*\H -0.25*\H *\L / \M+0.25*\H) circle (3pt);\draw(-\L,-2*\H -0.25*\H *\L / \M+0.25*\H) circle (3pt);
\filldraw[white!20!blue] (-\L,-2*\H +0.25*\H *\L / \M-0.25*\H) circle (3pt);\draw(-\L,-2*\H +0.25*\H *\L / \M-0.25*\H) circle (3pt);

\filldraw[white!20!blue] (\L,-2.75*\H -0.25*\H *\L / \M+0.25*\H) circle (3pt);\draw (\L,-2.75*\H -0.25*\H *\L / \M+0.25*\H)  circle (3pt);
\filldraw[white!20!blue] (\L,-2.75*\H +0.25*\H *\L / \M-0.25*\H) circle (3pt);\draw (\L,-2.75*\H+0.25*\H *\L / \M-0.25*\H)  circle (3pt);

\node[scale=2.0] at (0,-2.25*\H) { \textbf{$\vdots$}} ;

\node[scale=2.0] at (0,0-0.1*\H) { $\mathcal C_1$} ;
\node[scale=2.0] at (0,-0.5*\H-0.1*\H) { $\mathcal C_0$} ;
\node[scale=2.0] at (0.35,-\H-0.1*\H) { $\mathcal C_{-1}$} ;
\node[scale=2.0] at (0.35,-1.5*\H-0.1*\H) { $\mathcal C_{-2}$} ;
\node[scale=2.0] at (0.65,-2.75*\H-0.1*\H) { $\mathcal C_{-n^*}$} ;

\node[scale=2.0] at (\M-0.2, -0.5*\H+0.2*\H) { $\mathcal C_{1}^{-}$} ;
\node[scale=2.0] at (\M-0.2, -0.5*\H-0.2*\H) { $\mathcal C_{0}^{+}$} ;
\draw[line width=0.5mm, <->,>=stealth] (\M+0.35, -0.5*\H-0.1*\H) .. controls (\M+0.65,-0.5*\H)  .. (\M+0.35, -0.5*\H+0.1*\H);

\node[scale=2.0] at (\M-0.2, -1.5*\H+0.2*\H) { $\mathcal C_{-1}^{-}$} ;
\node[scale=2.0] at (\M-0.2, -1.5*\H-0.2*\H) { $\mathcal C_{-2}^{+}$} ;
\draw[line width=0.5mm, <->,>=stealth] (\M+0.35, -1.5*\H-0.1*\H) .. controls (\M+0.65,-1.5*\H)  .. (\M+0.35, -1.5*\H+0.1*\H);

\node[scale=2.0] at (\M-0.07, -2.75*\H+0.2*\H) { $\mathcal C_{-n^*\!+\!1}^{-}$} ;
\node[scale=2.0] at (\M-0.2, -2.75*\H-0.2*\H) { $\mathcal C_{-n^*}^{+}$} ;
\draw[line width=0.5mm, <->,>=stealth] (\M+0.35, -2.75*\H-0.1*\H-0.1) .. controls (\M+0.65,-2.75*\H-0.1)  .. (\M+0.35, -2.75*\H+0.1*\H-0.1);

\node[scale=2.0] at (-\M+0.2, -1*\H+0.2*\H) { $\mathcal C_{0}^{-}$} ;
\node[scale=2.0] at (-\M+0.2, -1*\H-0.2*\H) { $\mathcal C_{-1}^{+}$} ;
\draw[line width=0.5mm, <->,>=stealth] (-\M-0.3, -1*\H-0.1*\H) .. controls (-\M-0.6,-1*\H)  .. (-\M-0.3, -1*\H+0.1*\H);

\node[scale=2.0] at (-\M+0.2, -2*\H+0.2*\H) { $\mathcal C_{-2}^{-}$} ;
\node[scale=2.0] at (-\M+0.2, -2*\H-0.2*\H) { $\mathcal C_{-3}^{+}$} ;
\draw[line width=0.5mm, <->,>=stealth] (-\M-0.3, -2*\H-0.1*\H) .. controls (-\M-0.6,-2*\H)  .. (-\M-0.3, -2*\H+0.1*\H);

\node[scale=2.0] at (-\M+0.2, +0.2*\H) { $\mathcal C_{2}^{-}$} ;
\node[scale=2.0] at (-\M+0.2,-0.2*\H) { $\mathcal C_{1}^{+}$} ;
\draw[line width=0.5mm, <->,>=stealth] (-\M-0.3, -0.1*\H) .. controls (-\M-0.6,0)  .. (-\M-0.3, +0.1*\H);


\node[scale=2.0] at (0, -3.3*\H) { $Q[\mathcal C_{i>-n^*}]=\mathcal C_{-n^*}^-$} ;

\end{tikzpicture}}
\caption{The duality chain. Each straight line under $\mathcal C_i$ corresponds to the phase diagram of the theories under the common mass deformations of the matter fields. Blue regions are the semiclassical phases $\mathcal C_i^\pm$ obtained from the integrating out fermions with large positive/negative mass and are shared by asymptotic phase of adjacent theories. Each blue wavy line corresponds to the critical line with same universality class. As we go down the duality chain, quantum phase of $\mathcal C_0$ which denoted as $Q[\mathcal C_0]$ is determined by semiclassical phase of the $\mathcal C_{-n^*}$ within finite steps. Together with going up the duality chain, intermediate quantum phase is shared by infinitely many UV theories $\mathcal C_i$ with $i>-n^*$. } 
\end{figure}
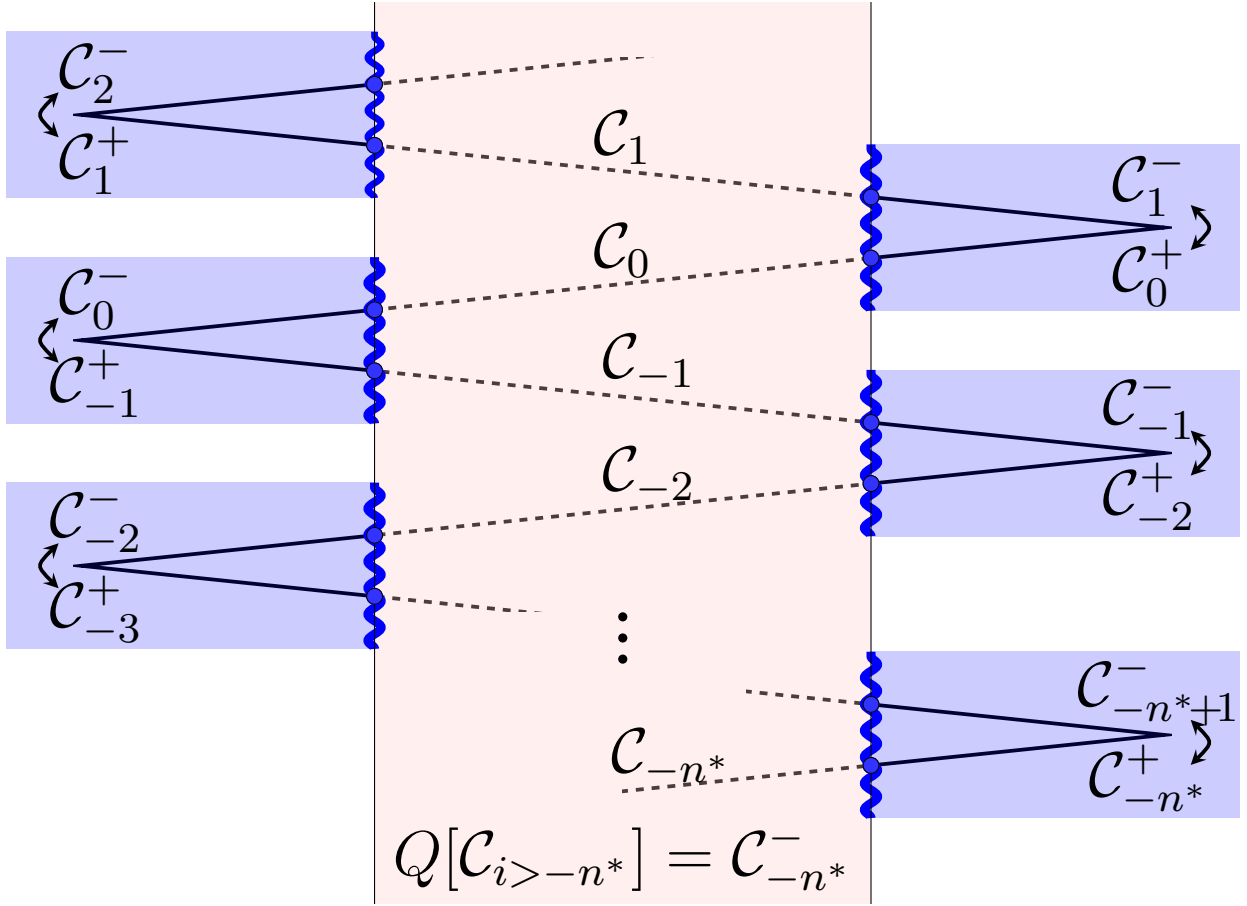 \label{fig:dualitychain}

\vspace{8pt}

{\begin{center}\underline{Outline of the Paper} \end{center}}

The outline of the paper is as follows. In section \ref{sec:review}, we review phase diagrams of single adjoint QCD$_3$ proposed in \cite{Gomis:2017ixy}. Section \ref{sec:phases} describes in detail the phase diagram and the duality chain for the case of 2 adjoint fermions under $SU/SO/Sp$ group and discuss its interesting dynamical implications. We extend the duality chain to the theories of two fermions in any combination of adjoint/rank-two representation or two bifundamental fermions in section \ref{sec:more dualities}. We present the various non-trivial consistency checks of the proposal using RG flows, generalized level-rank duality, Lie groups isomorphisms, $SL(2,\mathbb Z)$ transfromations and gravitational counterterms in section \ref{sec:additional_consistency_checks}. Finally, discussion and the possible future directions are in section \ref{sec:comments}. Appendix \ref{appsec:genlevelrank} explains the generalized level-rank duality of the unitary gauge group in terms of $SL(2,\mathbb Z)$ operations.

\section{Review : Phases of Single Adjoint QCD$_3$}\label{sec:review}

Before we propose and explain the phase diagram of two adjoints QCD$_3$ in the following section, we briefly review the proposal of single adjoint QCD$_3$ in \cite{Gomis:2017ixy}. Consider $G_k$ Yang-Mills-Chern-Simons theory with a gauge group $G$ and the CS level $k$ coupled to a real Majorana fermion $\lambda$ in the adjoint representation with its bare mass $m_\lambda$. For a special value of the bare mass $M_\lambda=m_{SUSY}\sim -kg^2$, theory exhibits the $\mathcal N=1$ supersymmetry and the index calculation in \cite{Witten:1999ds} shows that the supersymmetry is preserved for $k\geq h/2$ while predicts a spontaneous breaking of supersymmetry for $k<h/2$ with accompanying Goldstino in the IR. (h is the dual Coxeter number of the gauge group.)

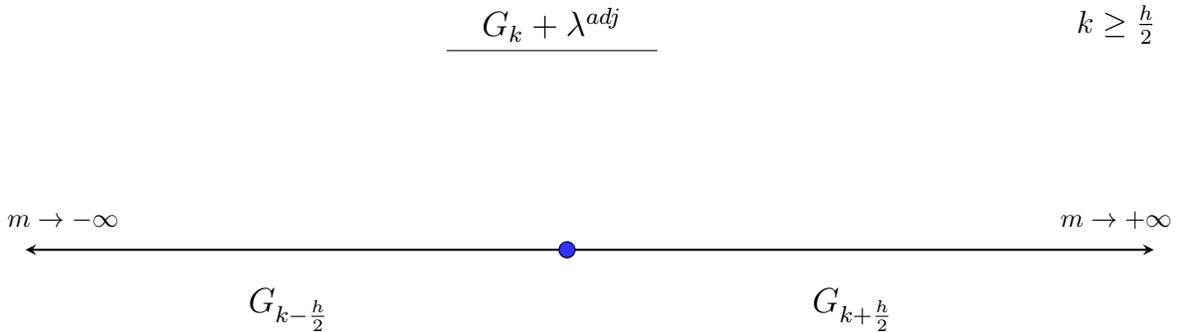
\begin{figure}[!h]
\centering
\begin{tikzpicture}
\node[scale=1.1] at (5.5,3) {$G_k+\lambda^{adj}$};
\node at (13,3) {$k\ge {h\ov2}$};
\draw (4.1,2.65) -- (6.9,2.65);
\draw[thick,<->,>=stealth] (-1.5,0) -- (13.5,0);
\filldraw[white!20!blue] (5.7,0) circle (3pt);\draw (5.7,0) circle (3pt);

\node[scale=1.1] at (2,-.8) {$G_{k-{h\ov2}}$};
\node[scale=1.1] at (9.5,-.8) {$G_{k+{h\ov 2}}$};

\node at (13,.4) {\footnotesize$m\to+\infty$};
\node at (-1,.4) {\footnotesize$m\to-\infty$};
\end{tikzpicture}
\caption{The proposal of \cite{Gomis:2017ixy} for the Phase diagram of $G_k$ gauge theory with a single real adjoint fermion for $k\geq h/2$. The solid circle represents a phase transition between the asymptotic phases.}
\label{fig:1adjGlargek}
\end{figure}

Let's first discuss the proposed phase diagram for $k\geq h/2$. Based on the large $k$ weakly-coupled analysis of Chern-Simons-matter theories \cite{Avdeev:1991za,Avdeev:1992jt}, the authors of \cite{Gomis:2017ixy} proposed the phase diagram for $k\geq h/2$ has a single phase transition separating two asymptotic Chern-Simons TQFTs as we depicted in the figure \ref{fig:1adjGlargek}. The two asymptotic TQFTs $G_{k \pm {h\ov2}}$ comes from the fact that integrating out the massive adjoint fermion shifts the Chern-Simons level in the IR by $\text{sgn}(m)\frac{h}{2}$ when we. 

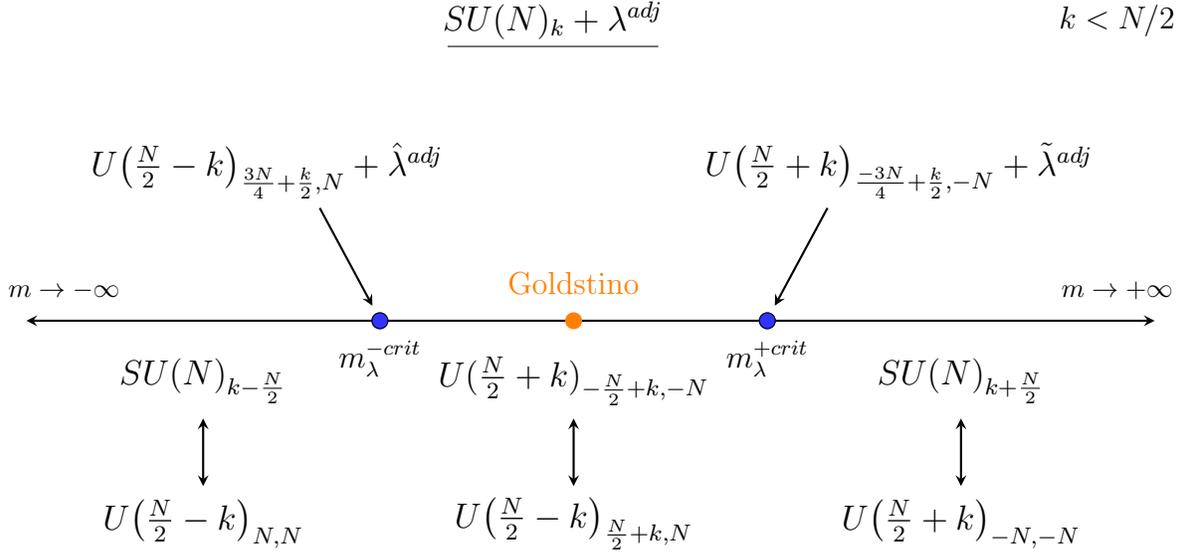
\begin{figure}[!h]
\centering
\begin{tikzpicture}
\node[scale=1.1] at (5.5,4) {$SU(N)_k+\lambda^{adj}$};
\node at (13,4) {$k<N/2$};
\draw (4.1,3.65) -- (6.9,3.65);
\draw[thick,<->,>=stealth] (-1.5,0) -- (13.5,0);
\filldraw[white!20!blue] (3.2,0) circle (3pt);\draw (3.2,0) circle (3pt);
\filldraw[white!20!blue] (8+.35,0) circle (3pt);\draw (8+.35,0) circle (3pt);
\node at (3.2,-0.5) {$ m_\lambda^{-crit}$};
\node at (8+.35,-0.5) {$ m_\lambda^{+crit}$};

\node[scale=1.1] at (1.7,2) {$ U\!\left({N\ov 2}-k\right)_{{3N\ov  4}+{k \ov 2},N}+\hat\lambda^{adj}$};
\node[scale=1.1] at (10.1,2) {$U\!\left({N\ov 2}+k\right)_{{-3N\ov  4}+{k \ov 2},-N}+\tilde\lambda^{adj}$};
\draw[thick,->,>=stealth] (2.4,1.5) -- (3.1,.2);
\draw[thick,->,>=stealth] (8.8+.35,1.5) -- (8.1+.35,.2);

\node[scale=1.1] at (.85,-.8) {$SU(N)_{k-{N\ov 2}}$};
\draw[thick,<->,>=stealth] (.85,-1.3) -- (.85,-2.2);
\node[scale=1.1] at (.85,-2.7) {$U\!\left({N\ov 2}-k\right)_{N,N}$};

\node[scale=1.1] at (5.7+.075,-.8) {$U({N\ov 2}+k)_{-{N\ov 2}+k,-N}$};
\draw[thick,<->,>=stealth] (5.7+.075,-1.3) -- (5.7+.075,-2.2);
\node[scale=1.1] at (5.7+.075,-2.7) {$U\!\left({N\ov 2}-k\right)_{{N\ov 2}+k,N}$};

\node at (5.775,0.5) {\textcolor{orange}{Goldstino}};
\filldraw[orange] (5.775,0) circle (3pt);

\node[scale=1.1] at (10.925,-.8) {$SU(N)_{k+{N\ov 2}}$};
\draw[thick,<->,>=stealth] (10.925,-1.3) -- (10.925,-2.2);
\node[scale=1.1] at (10.925,-2.7) {$U\!\left({N\ov 2}+k\right)_{-N,-N}$};

\node at (13,.4) {\footnotesize$m\to+\infty$};
\node at (-1,.4) {\footnotesize$m\to-\infty$};
\end{tikzpicture}
\caption{The proposal of \cite{Gomis:2017ixy} for the Phase diagram of $SU(N)_k$ gauge theory with a single real adjoint fermion for $k< N/2$. There are two phase transitions each described by a weakly-coupled dual gauge theory description, which appears with an arrow pointing to the phase transition. The intermediate phase from each dual descriptions are related by level/rank duality.}
\label{fig:1adjsu(n)smallk}
\end{figure}

For the case of $k<h/2$ (note that we chose $k\geq 0$ throught this paper.), it turns out that the IR phase of the supersymmetric point cannot be a single Goldstino alone due to the presence of the ’t Hooft anomalies. Let's focus on the case of $SU(N)$ gauge group. Most stringently, there is a $\mathbb Z_N$ 1-form symmetry which is anomalous \cite{Gaiotto:2014kfa} thus there must be some deconfined degrees of freedom in the IR. Remarkably, the authors of \cite{Gomis:2017ixy} proposed a consistent IR phase by thinking about the one-dimensional phase diagram in terms of bare mass $m_\lambda$ of the fermion. The natural prediction for the topology of the phase diagram is to have three distinct phases with two transition points, two of them are asymptotic semiclassical phases discussed above and the other one is intermediate non-perturbative phase where supersymmetry is broken. Now to preserve the $\mathbb Z_N$ 1-form symmetry of the original theory, the authors of \cite{Gomis:2017ixy} proposed a simple dual description at each transition point with a dual Yang-Mills-Chern-Simons theory coupled to a dual adjoint fermion. The impressive consequence of this two mutually non-local `weakly-coupled' dual descriptions is that they share a common intermediate phase by level/rank duality even though they were independently designed to share each positive or negative asymptotic phases of the original theory respectively. The phase diagram becomes figure \ref{fig:1adjsu(n)smallk} where the intermediate quantum phase is described by a TQFT with decoupled Goldstino. While the $\mathbb Z_N$ 1-form anomaly is automatically saturated by the construction, this phase diagram is also consistent with more subtle discrete time-reversal anomaly together with the background Riemannian metric \cite{Gomis:2017ixy}.

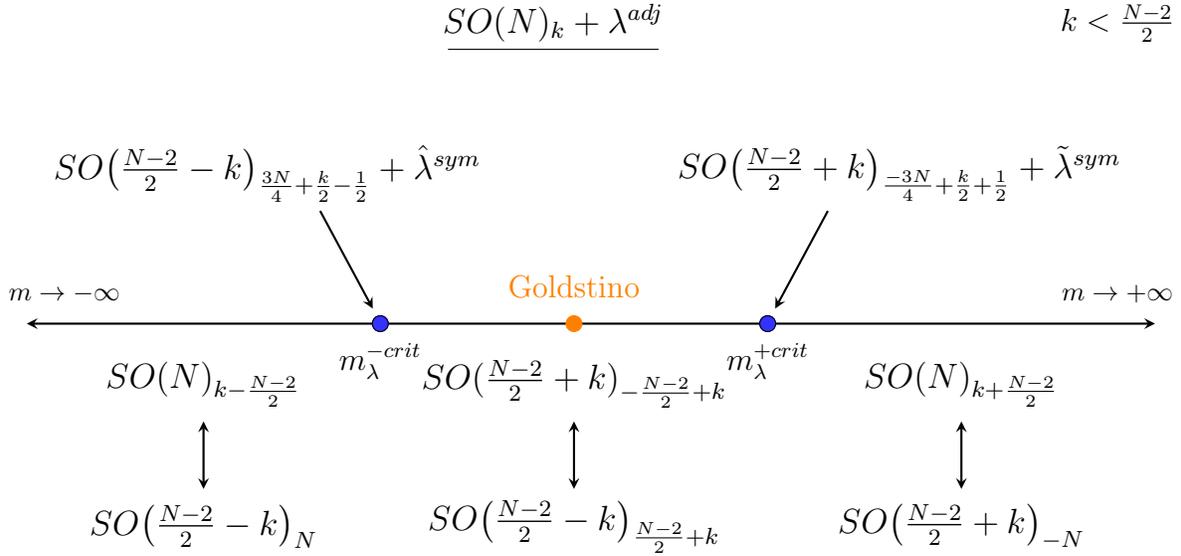
\begin{figure}[!h]
\centering
\begin{tikzpicture}
\node[scale=1.1] at (5.5,4) {$SO(N)_k+\lambda^{adj}$};
\node at (13,4) {$k<{N-2\ov 2}$};
\draw (4.1,3.65) -- (6.9,3.65);
\draw[thick,<->,>=stealth] (-1.5,0) -- (13.5,0);
\filldraw[white!20!blue] (3.2,0) circle (3pt);\draw (3.2,0) circle (3pt);
\filldraw[white!20!blue] (8+.35,0) circle (3pt);\draw (8+.35,0) circle (3pt);
\node at (3.2,-0.5) {$ m_\lambda^{-crit}$};
\node at (8+.35,-0.5) {$ m_\lambda^{+crit}$};
\node at (5.775,0.5) {\textcolor{orange}{Goldstino}};
\filldraw[orange] (5.775,0) circle (3pt);

\node[scale=1.1] at (1.7,2) {$ SO\!\left({N-2\ov 2}-k\right)_{{3N\ov  4}+{k \ov 2}-{1\ov 2}}+\hat\lambda^{sym}$};
\node[scale=1.1] at (10.1,2) {$SO\!\left({N-2\ov 2}+k\right)_{{-3N\ov  4}+{k \ov 2}+{1\ov 2}}+\tilde\lambda^{sym}$};
\draw[thick,->,>=stealth] (2.4,1.5) -- (3.1,.2);
\draw[thick,->,>=stealth] (8.8+.35,1.5) -- (8.1+.35,.2);

\node[scale=1.1] at (.85,-.8) {$SO(N)_{k-{N-2\ov 2}}$};
\draw[thick,<->,>=stealth] (.85,-1.3) -- (.85,-2.2);
\node[scale=1.1] at (.85,-2.7) {$SO\!\left({N-2\ov 2}-k\right)_{N}$};

\node[scale=1.1] at (5.7+.075,-.8) {$SO({N-2\ov 2}+k)_{-{N-2\ov 2}+k}$};
\draw[thick,<->,>=stealth] (5.7+.075,-1.3) -- (5.7+.075,-2.2);
\node[scale=1.1] at (5.7+.075,-2.7) {$SO\!\left({N-2\ov 2}-k\right)_{{N-2\ov 2}+k}$};

\node[scale=1.1] at (10.925,-.8) {$SO(N)_{k+{N-2\ov 2}}$};
\draw[thick,<->,>=stealth] (10.925,-1.3) -- (10.925,-2.2);
\node[scale=1.1] at (10.925,-2.7) {$SO\!\left({N-2\ov 2}+k\right)_{-N}$};

\node at (13,.4) {\footnotesize$m\to+\infty$};
\node at (-1,.4) {\footnotesize$m\to-\infty$};
\end{tikzpicture}
\caption{The proposal of \cite{Gomis:2017ixy} for the Phase diagram of $SO(N)_k$ gauge theory with a single real adjoint fermion for $k<{N-2 \ov 2}$. The representations of the dual fermions are changed to the other rank-two representation of the gauge group.}
\label{fig:1adjso(n)smallk}
\end{figure}

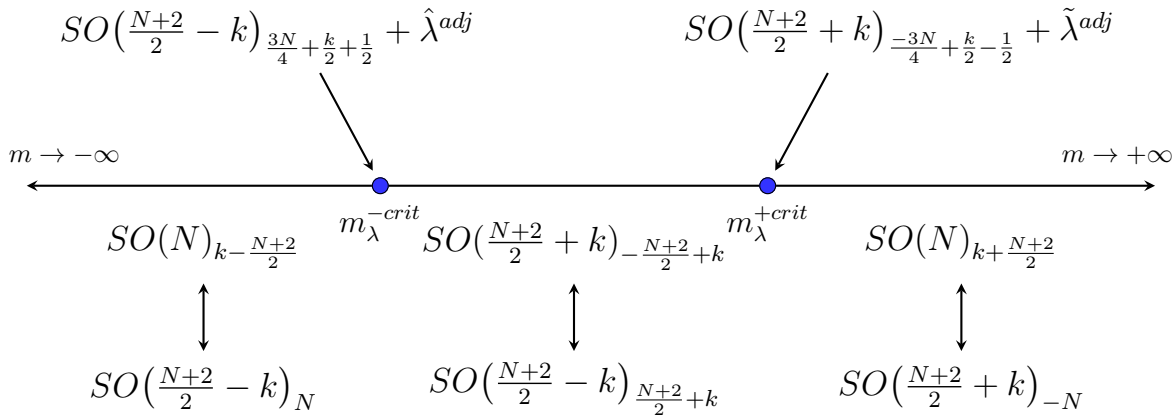
\begin{figure}[!h]
\centering
\begin{tikzpicture}
\node[scale=1.1] at (5.5,4) {$SO(N)_k+\lambda^{sym}$};
\node at (13,4) {$k<{N+2\ov 2}$};
\draw (4.1,3.65) -- (6.9,3.65);
\draw[thick,<->,>=stealth] (-1.5,0) -- (13.5,0);
\filldraw[white!20!blue] (3.2,0) circle (3pt);\draw (3.2,0) circle (3pt);
\filldraw[white!20!blue] (8+.35,0) circle (3pt);\draw (8+.35,0) circle (3pt);
\node at (3.2,-0.5) {$ m_\lambda^{-crit}$};
\node at (8+.35,-0.5) {$ m_\lambda^{+crit}$};

\node[scale=1.1] at (1.7,2) {$ SO\!\left({N+2\ov 2}-k\right)_{{3N\ov  4}+{k \ov 2}+{1\ov 2}}+\hat\lambda^{adj}$};
\node[scale=1.1] at (10.1,2) {$SO\!\left({N+2\ov 2}+k\right)_{{-3N\ov  4}+{k \ov 2}-{1\ov 2}}+\tilde\lambda^{adj}$};
\draw[thick,->,>=stealth] (2.4,1.5) -- (3.1,.2);
\draw[thick,->,>=stealth] (8.8+.35,1.5) -- (8.1+.35,.2);

\node[scale=1.1] at (.85,-.8) {$SO(N)_{k-{N+2\ov 2}}$};
\draw[thick,<->,>=stealth] (.85,-1.3) -- (.85,-2.2);
\node[scale=1.1] at (.85,-2.7) {$SO\!\left({N+2\ov 2}-k\right)_{N}$};

\node[scale=1.1] at (5.7+.075,-.8) {$SO({N+2\ov 2}+k)_{-{N+2\ov 2}+k}$};
\draw[thick,<->,>=stealth] (5.7+.075,-1.3) -- (5.7+.075,-2.2);
\node[scale=1.1] at (5.7+.075,-2.7) {$SO\!\left({N+2\ov 2}-k\right)_{{N+2\ov 2}+k}$};

\node[scale=1.1] at (10.925,-.8) {$SO(N)_{k+{N+2\ov 2}}$};
\draw[thick,<->,>=stealth] (10.925,-1.3) -- (10.925,-2.2);
\node[scale=1.1] at (10.925,-2.7) {$SO\!\left({N+2\ov 2}+k\right)_{-N}$};

\node at (13,.4) {\footnotesize$m\to+\infty$};
\node at (-1,.4) {\footnotesize$m\to-\infty$};
\end{tikzpicture}
\caption{The proposal of \cite{Gomis:2017ixy} for the Phase diagram of $SO(N)_k$ gauge theory with a single real symmetric-traceless fermion for $k<{N+2 \ov 2}$.}
\label{fig:1symso(n)smallk}
\end{figure}

Finally we comment on the case of $SO/Sp$ gauge group. The structure of the phase diagram for the $SO/Sp$ Yang-Mills-Chern-Simons theory with single real adjoint fermion has distinguished feature compared to the unitary group case. While the topology of the phase diagram still depends on whether $k\geq h/2$ or $k< h/2$, the dual fermions are no more in the adjoint representation of the dual gauge group. Instead, it is symmetric-traceless/antisymmetric-traceless (sym/asym) representation for the $SO/Sp$ gauge group to have a intermediate phase shared by level/rank duality. The 1-form symmetry is still preserved since the $SO/Sp$ gauge group has only trivial or $\mathbb Z_2$ one-form symmetry. Now, if we note that the adjoint representation of $SO/Sp$ is equivalent to the asym/sym representation, $SO/Sp$ gauge theory with single real fermion under sym/asym representation could be constructed in the same way, with dual descriptions containing single adjoint fermion. We summarized the $k<h/2$ phase diagram for the $SO/Sp$ gauge theory with a fermion in the rank-two representations in the figure \ref{fig:1adjso(n)smallk} and \ref{fig:1symso(n)smallk}.

\newpage

\section{Phase Diagrams for $k\neq 0$ : Duality Chain}
\label{sec:phases}

\subsection{$k\geq h$ : Semiclassical Regime}
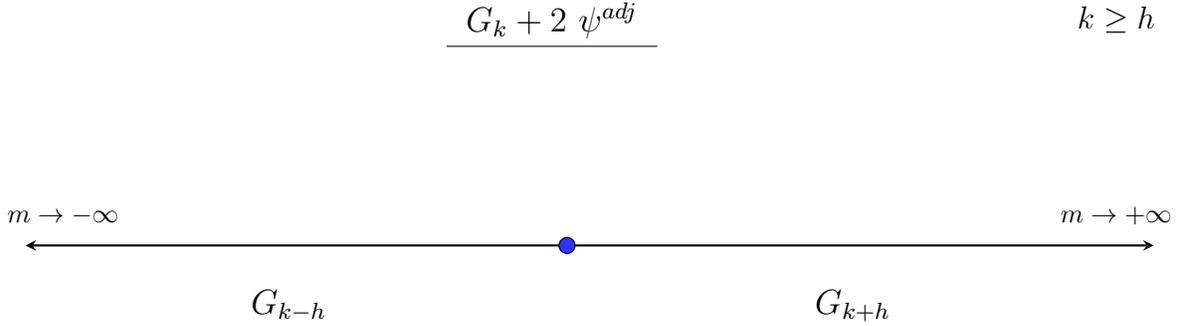
\begin{figure}[!h]
\centering
\begin{tikzpicture}
\node[scale=1.1] at (5.5,3) {$G_k+2~\psi^{adj}$};
\node at (13,3) {$k\ge h$};
\draw (4.1,2.65) -- (6.9,2.65);
\draw[thick,<->,>=stealth] (-1.5,0) -- (13.5,0);
\filldraw[white!20!blue] (5.7,0) circle (3pt);\draw (5.7,0) circle (3pt);

\node[scale=1.1] at (2,-.8) {$G_{k-h}$};
\node[scale=1.1] at (9.5,-.8) {$G_{k+h}$};

\node at (13,.4) {\footnotesize$m\to+\infty$};
\node at (-1,.4) {\footnotesize$m\to-\infty$};
\end{tikzpicture}
\caption{Phase diagram of $G_k$ gauge theory with two real adjoint fermions for $k\geq h$. The solid circle represents a phase transition between the asymptotic phases. For sufficiently large $k$ we know for certain \cite{Avdeev:1991za,Avdeev:1992jt} that the phase transition is associated with a CFT.}
\label{largek}
\end{figure}

Now we consider the Yang-Mills-Chern-Simons theory with the gauge group $G$ and the effective Chern-Simons level $k$ coupled with two real adjoint fermions. Similar to the single adjoint phase diagram, as long as $k$ is sufficiently large, the above two different topological phases $G_{k+\sign(m) h}$ are separated by a single transition \cite{Avdeev:1991za, Avdeev:1992jt}. We call this as an semiclassical phase diagram. Then the main question is at which value of $k$ topology of the phase diagram changes and additional phases appear. We propose that to have a consistent picture, the topology of large $k$ phase diagram should hold for $k\geq k_{crit}=h$. (Recall that we choose $k\geq 0$ from the beginning.) The phase diagram for $k\geq h$ is given in the figure \ref{largek}.

\subsection{Quantum Phase for $G=SU(N)$}

For $0< k<h$, we propose that there is a new single intermediate ``quantum phase" in between the asymptotic large mass phases, which we denote as $Q[G_k^R]$ where $G$ is dynamical gauge group and $k$ is UV Chern-Simons level together with matter fields in the representation $R$. This inherently non-perturbative phase is connected to the two asymptotic phases at infinity through the phase transitions at $m= m_\psi^{\pm crit}$. Now to access the intermediate phase, we need to find a consistent proposal for the dual descriptions describing each critical points. Motivated from the fermion-fermion dualities for the single adjoint fermion case we reviewed in the section \ref{sec:review}, we propose that each dual description is described by a dual Yang-Mills-Chern-Simons theory coupled to two dual adjoint fermions. Explicitly, we suggest the following mutually non-local dual descriptions of the original theory at quantum phase regime as follows :

 \smallskip
\noindent Dualities for $SU(N)_k+~\text{2}~ \psi^{adj}$ for $0<k<N$:
\begin{equation} \label{eq:su(n)smallk}
\begin{aligned}
&SU(N)_k+~\text{2}~ \psi^{adj},~ m_\psi=m^{+crit}_\psi &\longleftrightarrow~~ &U\!\left(N+k\right)_{k,-N}+~\text{2}~ \tilde \psi^{adj},~ m_{\tilde \psi}=m^{-crit}_{\tilde \psi} \\[+5pt]
&SU(N)_k+~\text{2}~ \psi^{adj},~ m_{\psi}=m^{-crit}_{\psi} &\longleftrightarrow~~ &U\!\left(N-k\right)_{k,N}+~\text{2}~\hat \psi^{adj},~ m_{\hat \psi}=m^{+crit}_{\hat \psi}\,.
\end{aligned}
\end{equation}

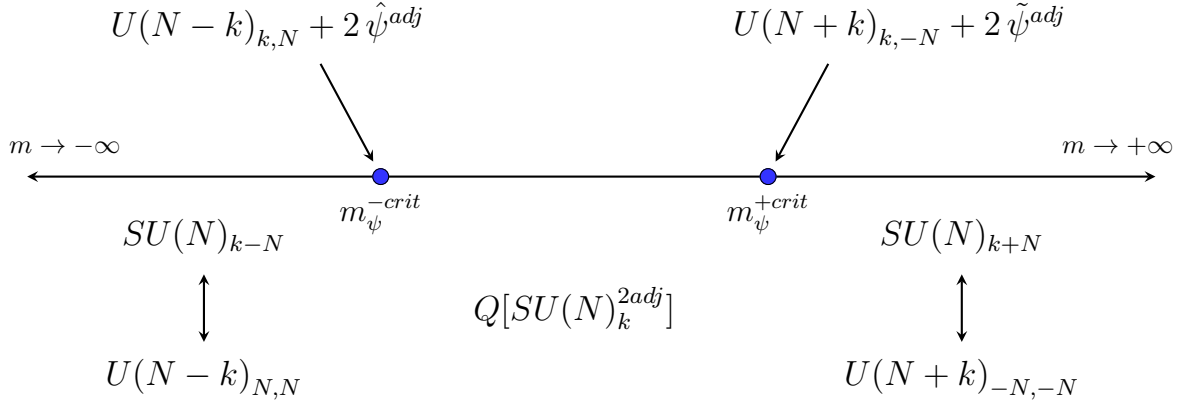
\begin{figure}[!h]
\centering
\begin{tikzpicture}
\node[scale=1.1] at (5.5,4) {$SU(N)_k+2\,\psi^{adj}$};
\node at (13,4) {$0<k<N$};
\draw (4.1,3.65) -- (6.9,3.65);
\draw[thick,<->,>=stealth] (-1.5,0) -- (13.5,0);
\filldraw[white!20!blue] (3.2,0) circle (3pt);\draw (3.2,0) circle (3pt);
\filldraw[white!20!blue] (8+.35,0) circle (3pt);\draw (8+.35,0) circle (3pt);
\node at (3.2,-0.5) {$ m_\psi^{-crit}$};
\node at (8+.35,-0.5) {$ m_\psi^{+crit}$};

\node[scale=1.1] at (1.7,2) {$ U\!\left(N-k\right)_{k,N}+2\,\hat\psi^{adj}$};
\node[scale=1.1] at (10.1,2) {$U\!\left(N+k\right)_{k,-N}+2\,\tilde\psi^{adj}$};
\draw[thick,->,>=stealth] (2.4,1.5) -- (3.1,.2);
\draw[thick,->,>=stealth] (8.8+.35,1.5) -- (8.1+.35,.2);

\node[scale=1.1] at (.85,-.8) {$SU(N)_{k-N}$};
\draw[thick,<->,>=stealth] (.85,-1.3) -- (.85,-2.2);
\node[scale=1.1] at (.85,-2.7) {$U\!\left(N-k\right)_{N,N}$};

\node[scale=1.1] at (5.7+.075,-1.75) {$Q[SU(N)_k^{2adj}]$};
\node[scale=1.1] at (10.925,-.8) {$SU(N)_{k+N}$};
\draw[thick,<->,>=stealth] (10.925,-1.3) -- (10.925,-2.2);
\node[scale=1.1] at (10.925,-2.7) {$U\!\left(N+k\right)_{-N,-N}$};

\node at (13,.4) {\footnotesize$m\to+\infty$};
\node at (-1,.4) {\footnotesize$m\to-\infty$};
\end{tikzpicture}
\caption{Phase diagram of $SU(N)$ with two real adjoint fermions for $0<k<N$. The solid circles represent a phase transition between the asymptotic phases and the intermediate quantum phase. Each phase transition has a dual gauge theory description, which appears with an arrow pointing to the phase transition. The mass deformations are related by 
$\delta m_\psi=-\delta m_{\hat\psi}$ and $\delta m_\psi=-\delta m_{\tilde\psi}$. Importantly, the right dual description is always strongly coupled and left dual description is weakly coupled only when $N/2 \leq k<N$. }
\label{fig:su(n)smallk}
\end{figure}

 Here, the distinguishing feature compared to the single adjoint phase diagram reviewed in the section \ref{sec:review} comes from our assumption of $k_{crit}=h$\footnote{It is important to note that condition for the quantum phase for the case of $U(N)_{k,k'}$ UV gauge group is same as $SU(N)$ case, i.e. $k_{crit}=h$ and abelian level doesn't play any role. This is due to the fact that we could gauge the U(1) global baryon/monopole symmetry in the phase diagram to get another consistent phase diagram differing only by the content of abelian gauge group and its Chern-Simons level in the UV.}. From the consistency, we immediately see that the dual description at $m=m_\psi^{+crit}$ is always strongly coupled and the other dual description at $m=m_\psi^{-crit}$ is weakly coupled only when $k\geq N/2$. Since the dual theory is generically strongly coupled too, which has two transition points at $ m^{\pm crit}$, it was necessary to carefully specify the dual theory with the location of bare mass in the \eqref{eq:su(n)smallk}. The phase diagram is depicted in figure \ref{fig:su(n)smallk}.

Before introducing a \emph{`Duality Chain'}, it is vital to note that the assumption of $k_{crit}=h$ makes the left dual description at $m_\psi=m_\psi^{-crit}$ weakly-coupled for $N/2 \leq k <N$. Thus the quantum phase of $Q[SU(N)_k^{2adj}]$ is directly obtainable in this range as $Q[SU(N)_k^{2adj}]=U(N-k)_{2k-N,N}$.

\medskip
\noindent
$\bullet$ {\it$0<k<N/2$ : Duality Chain}
\medskip

\begin{figure}[!h]
\centering
\begin{tikzpicture}
\node[scale=1.1] at (5.5,4) {$U(N-k)_{k,N}+2\,\hat \psi^{adj}$};
\node at (13,4) {$0<k<N/2$};
\draw (4.1,3.65) -- (6.9,3.65);
\draw[thick,<->,>=stealth] (-1.5,0) -- (13.5,0);
\filldraw[white!20!blue] (3.2,0) circle (3pt);\draw (3.2,0) circle (3pt);
\filldraw[white!20!blue] (8+.35,0) circle (3pt);\draw (8+.35,0) circle (3pt);
\node at (3.2,-0.5) {$ -m_{\hat \psi}^{crit}$};
\node at (8+.35,-0.5) {$ m_{\hat \psi}^{crit}$};

\node[scale=1.1] at (1.7,2) {$ ST^2U\!\left(N-2k\right)_{k,N-k}+2\,\bar \psi^{adj}$};
\node[scale=1.1] at (10.1,2) {$SU\!\left(N\right)_{k}+2\,\psi^{adj}$};
\draw[thick,->,>=stealth] (2.4,1.5) -- (3.1,.2);
\draw[thick,->,>=stealth] (8.8+.35,1.5) -- (8.1+.35,.2);

\node[scale=1.1] at (.85,-.8) {$U\!\left(N-k\right)_{2k-N,N}$};
\draw[thick,<->,>=stealth] (.85,-1.3) -- (.85,-2.2);
\node[scale=1.1] at (.85,-2.7) {$ST^2 U\!\left(N-2k \right)_{N-k,N-k}$};

\node[scale=1.1] at (5.7+.075,-1.5) {$Q[U(N-k)_{k, N}^{2adj}]$};
\node[scale=1.1] at (5.7+.075,-2.25) {$=Q[SU(N)_k^{2adj}]$};
\node[scale=1.1] at (10.925,-.8) {$U\!\left(N-k\right)_{N,N}$};
\draw[thick,<->,>=stealth] (10.925,-1.3) -- (10.925,-2.2);
\node[scale=1.1] at (10.925,-2.7) {$SU(N)_{k-N}$};

\node at (13,.4) {\footnotesize$m\to+\infty$};
\node at (-1,.4) {\footnotesize$m\to-\infty$};
\end{tikzpicture}
\caption{Phase diagram of strongly coupled left dual description of figure \ref{fig:su(n)smallk}, which is $U(N-k)_{k,N}$ with two real adjoint fermions for $0<k<N/2$. It describes the first step of duality chain, where the right dual description at $m_{\hat \psi}=m_{\hat \psi}^{+crit}$ coincides with original theory thus shares the same quantum phase with original theory as $Q[SU(N)_k^{2adj}]=Q[U(N-k)_{k,N}]$. Left dual description has lower rank with same Chern-Simons level, thus one could use this chain recursively to get weakly coupled description of quantum phase within finite steps. The mass deformations are related by 
$\delta m_{\hat \psi}=- \delta m_{\psi}$ and $\delta m_{\hat \psi}=-\delta m_{\bar \psi}$.}
\label{fig:su(n)firstdualitychain}
\end{figure}
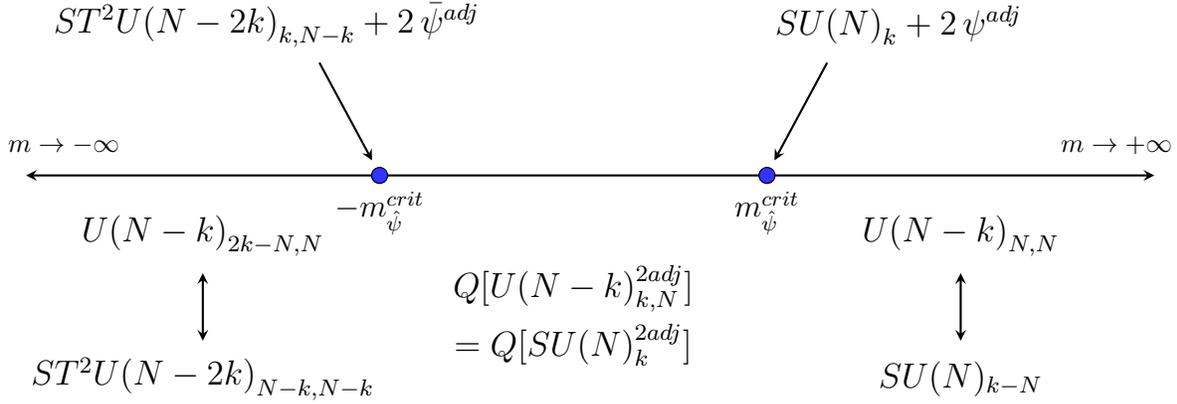

Now for the case of $0<k<N/2$, both dual descriptions in the figure \ref{fig:su(n)smallk} are strongly coupled. Remarkably, we can nevertheless obtain the description of the quantum phase $Q[SU(N)_k^{2adj}]$ only from the semiclasscal analysis. Let's first draw the phase diagram of the dual description at $m_\psi=m_\psi^{-crit}$. Since we are in the range of $0<k<N/2$, the left dual description is still strongly coupled and we could draw its own phase diagram as in the figure \ref{fig:su(n)firstdualitychain}.

Now things become transparent when look at the effective Chern-Simons level and rank of various UV descriptions in the phase diagram of original theory and left dual theory in the figure \ref{fig:su(n)smallk} and \ref{fig:su(n)firstdualitychain}. The observation is that in each of the phase diagram the left dual description at $m=m^{-crit}$ has smaller rank than original gauge group by $k$ with the same UV Chern-Simons level $k$. Hence for the non zero $k$, we could use these steps repeatedly to finally get the weakly coupled dual description which shares the same quantum phase $Q[SU(N)_k^{2adj}]$ of original theory. We dub this as \emph{`Duality Chain'} because we are keep using a different mutually non-local dual descriptions of the strongly coupled theory to reach the final weakly coupled description. Once we arrive at the final stage of a duality chain, we find a weakly coupled description flows to the original quantum phase which could be determined by the semiclassical analysis. Importantly, after first step of the chain, we need to use the generalized level-rank duality described in the appendix \ref{appsec:genlevelrank}. We further emphasize that critical point of the final weakly coupled dual description is not dual (for $0<k<N/2$) to any of the original strongly coupled theory's critical point.

\begin{tcolorbox}
Using the duality chain, we get the general expression for the quantum phase of $SU(N)_k+2~\psi^{adj}$ for $0<k<N$ which is identified after $\ceil{\frac{N}{k}}-1$ steps :

\begin{empheq}[box=\fbox]{equation}
\label{eq:su(n)qphase}
Q[SU(N)_k^{2adj}]= (ST^2)^{\ceil{\frac{N}{k}}-2}U\!\left(N-\ceil{\frac{N}{k}}k+k\right)_{-N+\ceil{\frac{N}{k}}k,N-\ceil{\frac{N}{k}}k+2k}.
\end{empheq}

\end{tcolorbox}

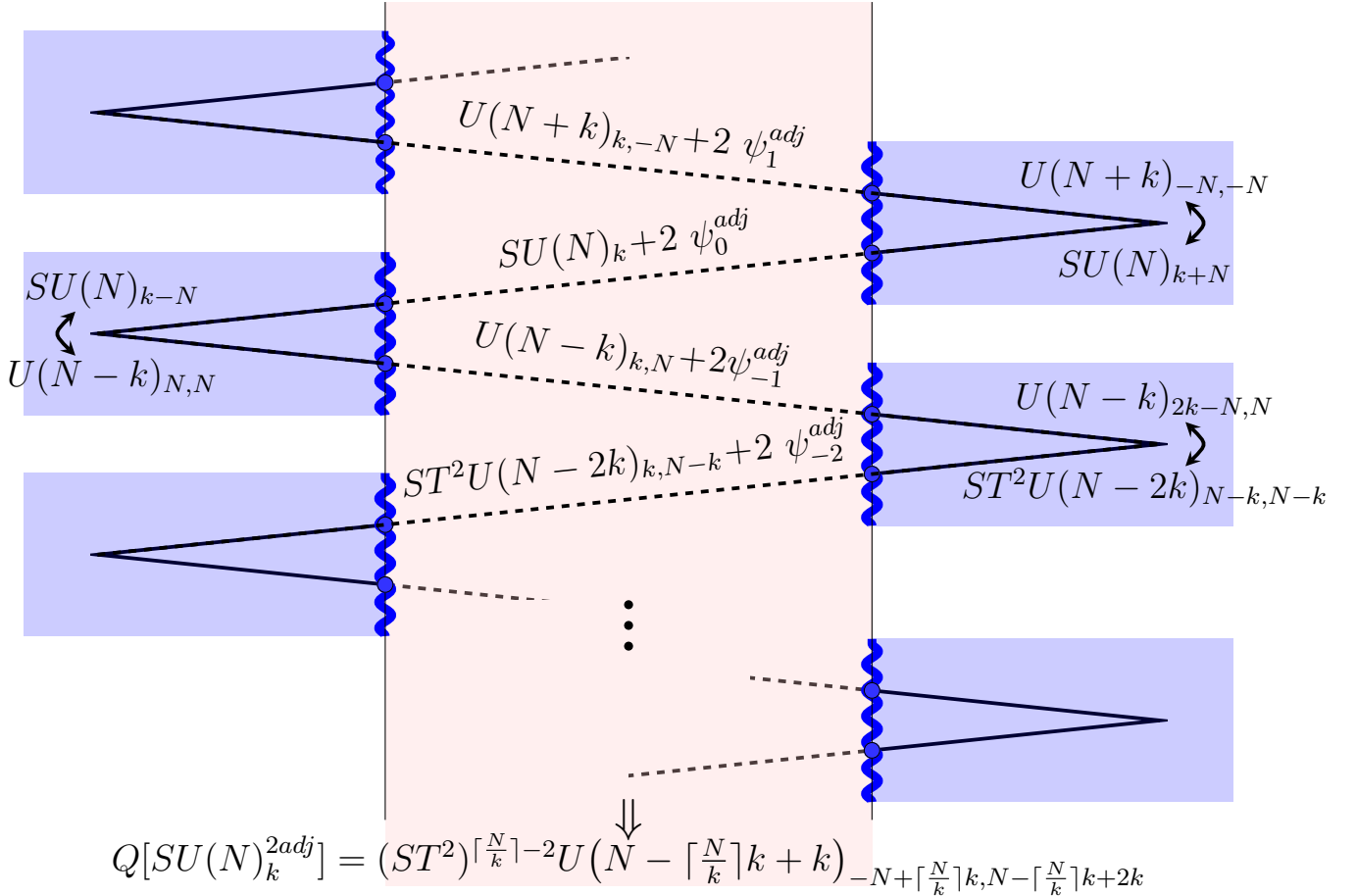
\begin{figure} [!h]
\makebox[\textwidth][c]{
\begin{tikzpicture}
\pgfmathsetmacro{\M}{7.2}
\pgfmathsetmacro{\L}{3.3}
\pgfmathsetmacro{\H}{3.0}


\begin{scope}
\clip (\L,-2*\H-.1) rectangle (\M+.1,.1);
\draw[thick,->,>=stealth,line width=0.5mm] (-\M,0) -- (\M,-0.5*\H) -- (-\M,-\H)  -- (\M,-1.5*\H)-- (-\M,-2*\H) -- (\M,-2.5*\H);
\end{scope}

\begin{scope}
\clip (-\L,-3*\H-.1) rectangle (-\M-.1,\H+.1);
\draw[black, thick,line width=0.5mm]  (\M,0.5*\H) -- (-\M,0) -- (\M,-0.5*\H) -- (-\M,-\H)  -- (\M,-1.5*\H)-- (-\M,-2*\H) --(\M, -2.5*\H );
\end{scope}

\begin{scope}
\clip (-\L,-2.17*\H-.1) rectangle (0,-2*\H+.1);
\draw[dashed,thick,line width=0.5mm] (\M,0.5*\H) --  (-\M,0) -- (\M,-0.5*\H) -- (-\M,-\H)  -- (\M,-1.5*\H)-- (-\M,-2*\H) -- (\M,-2.5*\H);
\end{scope}

\begin{scope}
\clip (-\L,0) rectangle (0,0.25*\H);
\draw[dashed,thick,line width=0.5mm] (\M,0.5*\H) --  (-\M,0) ;
\end{scope}

\begin{scope}
\clip (\L,-4*\H-.1) rectangle (\M+.1,.1);
\draw[thick,->,>=stealth,line width=0.5mm,black] (-\M,-2.25*\H)--(\M,-2.75*\H)--(-\M, -3.25*\H);
\end{scope}

\begin{scope}
\clip (0.5*\L,-3*\H-.1) rectangle (\L,.1);
\draw[dashed,thick,line width=0.5mm,black] (-\M,-2.25*\H)--(\M,-2.75*\H);
\end{scope}

\begin{scope}
\clip (0,-3*\H-.1) rectangle (\L,.1);
\draw[dashed,thick,line width=0.5mm,black] (\M,-2.75*\H)--(-\M, -3.25*\H);
\end{scope}

\fill[red!20!white, fill opacity=0.3]  (-\L, -3.5*\H) rectangle  (\L, 0.5*\H) ;

\fill[blue, fill opacity=0.2]  (-\M-1,+0.25*\H *\L / \M-0.25*\H -0.7) rectangle  (-\L,-0.25*\H *\L / \M+0.25*\H +0.7) ;
\draw[blue,decorate, decoration={snake},line width=0.9mm,opacity=1] (-\L,+0.25*\H *\L / \M-0.25*\H -0.7)--(-\L,-0.25*\H *\L / \M+0.25*\H +0.7);

\fill[blue, fill opacity=0.2]  (-\M-1,-\H+0.25*\H *\L / \M-0.25*\H -0.7) rectangle  (-\L,-\H-0.25*\H *\L / \M+0.25*\H +0.7) ;
\draw[blue,decorate, decoration={snake},line width=1.2mm,opacity=1] (-\L,-\H+0.25*\H *\L / \M-0.25*\H -0.7)--(-\L,-\H-0.25*\H *\L / \M+0.25*\H +0.7);

\fill[blue, fill opacity=0.2]  (-\M-1,-2*\H+0.25*\H *\L / \M-0.25*\H -0.7) rectangle  (-\L,-2*\H-0.25*\H *\L / \M+0.25*\H +0.7) ;
\draw[blue,decorate, decoration={snake},line width=1.2mm,opacity=1] (-\L,-2*\H+0.25*\H *\L / \M-0.25*\H -0.7)--(-\L,-2*\H-0.25*\H *\L / \M+0.25*\H +0.7);

\fill[blue, fill opacity=0.2]   (\L,-0.5*\H+0.25*\H *\L / \M-0.25*\H -0.7) rectangle  (\M+1,-0.5*\H-0.25*\H *\L / \M+0.25*\H +0.7) ;
\draw[blue,decorate, decoration={snake},line width=1.2mm,opacity=1] (\L,-0.5*\H+0.25*\H *\L / \M-0.25*\H -0.7)--(\L,-0.5*\H-0.25*\H *\L / \M+0.25*\H +0.7);

\fill[blue, fill opacity=0.2]   (\L,-1.5*\H+0.25*\H *\L / \M-0.25*\H -0.7) rectangle  (\M+1,-1.5*\H-0.25*\H *\L / \M+0.25*\H +0.7) ;
\draw[blue,decorate, decoration={snake},line width=1.2mm,opacity=1] (\L,-1.5*\H+0.25*\H *\L / \M-0.25*\H -0.7)--(\L,-1.5*\H-0.25*\H *\L / \M+0.25*\H +0.7);

\fill[blue, fill opacity=0.2]   (\L,-2.75*\H+0.25*\H *\L / \M-0.25*\H -0.7) rectangle  (\M+1,-2.75*\H-0.25*\H *\L / \M+0.25*\H +0.7) ;
\draw[blue,decorate, decoration={snake},line width=1.2mm,opacity=1] (\L,-2.75*\H+0.25*\H *\L / \M-0.25*\H -0.7)--(\L,-2.75*\H-0.25*\H *\L / \M+0.25*\H +0.7);

\draw (-\L, 0.5*\H) -- (-\L, -3.2*\H);
\draw (\L, 0.5*\H) -- (\L, -3.2*\H);

\filldraw[white!20!blue] (-\L,0-0.25*\H *\L / \M+0.25*\H) circle (3pt);\draw (-\L,0-0.25*\H *\L / \M+0.25*\H) circle (3pt);
\filldraw[white!20!blue] (-\L,0+0.25*\H *\L / \M-0.25*\H) circle (3pt);\draw (-\L,0+0.25*\H *\L / \M-0.25*\H) circle (3pt);

\filldraw[white!20!blue] (\L,-0.5*\H -0.25*\H *\L / \M+0.25*\H) circle (3pt);\draw (\L,-0.5*\H -0.25*\H *\L / \M+0.25*\H)  circle (3pt);
\filldraw[white!20!blue] (\L,-0.5*\H +0.25*\H *\L / \M-0.25*\H) circle (3pt);\draw (\L,-0.5*\H+0.25*\H *\L / \M-0.25*\H)  circle (3pt);

\filldraw[white!20!blue] (-\L,-\H -0.25*\H *\L / \M+0.25*\H) circle (3pt);\draw(-\L,-\H -0.25*\H *\L / \M+0.25*\H) circle (3pt);
\filldraw[white!20!blue] (-\L,-\H +0.25*\H *\L / \M-0.25*\H) circle (3pt);\draw(-\L,-\H +0.25*\H *\L / \M-0.25*\H) circle (3pt);

\filldraw[white!20!blue] (\L,-1.5*\H -0.25*\H *\L / \M+0.25*\H) circle (3pt);\draw (\L,-1.5*\H -0.25*\H *\L / \M+0.25*\H)  circle (3pt);
\filldraw[white!20!blue] (\L,-1.5*\H +0.25*\H *\L / \M-0.25*\H) circle (3pt);\draw (\L,-1.5*\H+0.25*\H *\L / \M-0.25*\H)  circle (3pt);

\filldraw[white!20!blue] (-\L,-2*\H -0.25*\H *\L / \M+0.25*\H) circle (3pt);\draw(-\L,-2*\H -0.25*\H *\L / \M+0.25*\H) circle (3pt);
\filldraw[white!20!blue] (-\L,-2*\H +0.25*\H *\L / \M-0.25*\H) circle (3pt);\draw(-\L,-2*\H +0.25*\H *\L / \M-0.25*\H) circle (3pt);

\filldraw[white!20!blue] (\L,-2.75*\H -0.25*\H *\L / \M+0.25*\H) circle (3pt);\draw (\L,-2.75*\H -0.25*\H *\L / \M+0.25*\H)  circle (3pt);
\filldraw[white!20!blue] (\L,-2.75*\H +0.25*\H *\L / \M-0.25*\H) circle (3pt);\draw (\L,-2.75*\H+0.25*\H *\L / \M-0.25*\H)  circle (3pt);

\node[scale=2.0] at (0,-2.25*\H) { \textbf{$\vdots$}} ;

\draw[dashed,thick,line width=0.5mm] (-\M,-0*\H) -- (\M,-0.5*\H) node[midway, above, sloped,scale=1.2] (TextNode) { $U(N+k)_{k,-N}\!+\!2~\psi_1^{adj}$} ;
\draw[dashed,thick,line width=0.5mm] (-\M,-\H) -- (\M,-0.5*\H) node[midway, above, sloped,scale=1.2] (TextNode) { $SU(N)_k\!+\!2~\psi_0^{adj}$} ;
\draw[dashed,thick,line width=0.5mm] (-\M,-\H) -- (\M,-1.5*\H) node[midway, above, sloped,scale=1.2] (TextNode) { $U(N-k)_{k,N}\!+\!2\psi_{-1}^{adj}$} ;
\draw[dashed,thick,line width=0.5mm] (-\M,-2*\H) -- (\M,-1.5*\H) node[midway, above, sloped,scale=1.2] (TextNode) { $ST^{2}U(N-2k)_{k,N-k}\!+\!2~\psi_{-2}^{adj}$} ;

\node[scale=1.2] at (\M-0.2, -0.5*\H+0.2*\H) { $U(N+k)_{-N,-N}$} ;
\node[scale=1.2] at (\M-0.2, -0.5*\H-0.2*\H) { $ SU(N)_{k+N}$} ;
\draw[line width=0.5mm, <->,>=stealth] (\M+0.35, -0.5*\H-0.1*\H) .. controls (\M+0.65,-0.5*\H)  .. (\M+0.35, -0.5*\H+0.1*\H);

\node[scale=1.2] at (\M-0.2, -1.5*\H+0.2*\H) { $ U(N-k)_{2k-N,N}$} ;
\node[scale=1.2] at (\M-0.2, -1.5*\H-0.2*\H) { $ ST^2U(N-2k)_{N-k,N-k}$} ;
\draw[line width=0.5mm, <->,>=stealth] (\M+0.35, -1.5*\H-0.1*\H) .. controls (\M+0.65,-1.5*\H)  .. (\M+0.35, -1.5*\H+0.1*\H);


\node[scale=1.2] at (-\M+0.2, -1*\H+0.2*\H) { $SU(N)_{k-N}$} ;
\node[scale=1.2] at (-\M+0.2, -1*\H-0.2*\H) { $U(N-k)_{N,N}$} ;
\draw[line width=0.5mm, <->,>=stealth] (-\M-0.3, -1*\H-0.1*\H) .. controls (-\M-0.6,-1*\H)  .. (-\M-0.3, -1*\H+0.1*\H);

\node[scale=1.	5] at (0,-3.2*\H) {$\Downarrow$};




\node[scale=1.2] at (0, -3.4*\H) { $Q[SU(N)_k^{2adj}]= (ST^2)^{\ceil{\frac{N}{k}}-2}U\!\left(N-\ceil{\frac{N}{k}}k+k\right)_{-N+\ceil{\frac{N}{k}}k,N-\ceil{\frac{N}{k}}k+2k}$} ;
-

\end{tikzpicture}}
\caption{The duality chain for the theory of $SU(N)_k+2~\psi^{adj}$.}
 \label{fig:dualitychainSU}  \end{figure}

In summary, global picture of duality chain for the case of $SU(N)_k+2~\psi^{adj}$ is illustrated in the figure \ref{fig:dualitychainSU}

\subsection{Quantum Phase for $G=SO(N)$}

\begin{figure}[!h]
\centering
\begin{tikzpicture}
\node[scale=1.1] at (5.5,4) {$SO(N)_k+2\,\psi^{adj}$};
\node at (13,4) {$0<k<N-2$};
\draw (4.1,3.65) -- (6.9,3.65);
\draw[thick,<->,>=stealth] (-1.5,0) -- (13.5,0);
\filldraw[white!20!blue] (3.2,0) circle (3pt);\draw (3.2,0) circle (3pt);
\filldraw[white!20!blue] (8+.35,0) circle (3pt);\draw (8+.35,0) circle (3pt);
\node at (3.2,-0.5) {$ m_\psi^{-crit}$};
\node at (8+.35,-0.5) {$ m_\psi^{+crit}$};

\node[scale=1.1] at (1.7,2) {$ SO\!\left(N-k-2\right)_{k}+2\,\hat\psi^{sym}$};
\node[scale=1.1] at (10.1,2) {$SO\!\left(N+k-2\right)_{k}+2\,\tilde\psi^{sym}$};
\draw[thick,->,>=stealth] (2.4,1.5) -- (3.1,.2);
\draw[thick,->,>=stealth] (8.8+.35,1.5) -- (8.1+.35,.2);

\node[scale=1.1] at (.85,-.8) {$SO(N)_{k-N+2}$};
\draw[thick,<->,>=stealth] (.85,-1.3) -- (.85,-2.2);
\node[scale=1.1] at (.85,-2.7) {$SO\!\left(N-k-2\right)_{N}$};

\node[scale=1.1] at (5.7+.075,-1.75) {$Q[SO(N)_k^{2adj}]$};
\node[scale=1.1] at (10.925,-.8) {$SO(N)_{k+N-2}$};
\draw[thick,<->,>=stealth] (10.925,-1.3) -- (10.925,-2.2);
\node[scale=1.1] at (10.925,-2.7) {$SO\!\left(N+k-2\right)_{-N}$};

\node at (13,.4) {\footnotesize$m\to+\infty$};
\node at (-1,.4) {\footnotesize$m\to-\infty$};
\end{tikzpicture}
\caption{Phase diagram of $SO(N)$ with two real adjoint fermions for $0<k<N-2$. The right dual description is always strongly coupled and left dual description is weakly coupled only when $N/2 \leq k<N-2$. }
\label{fig:so(n)adjsmallk}
\end{figure}
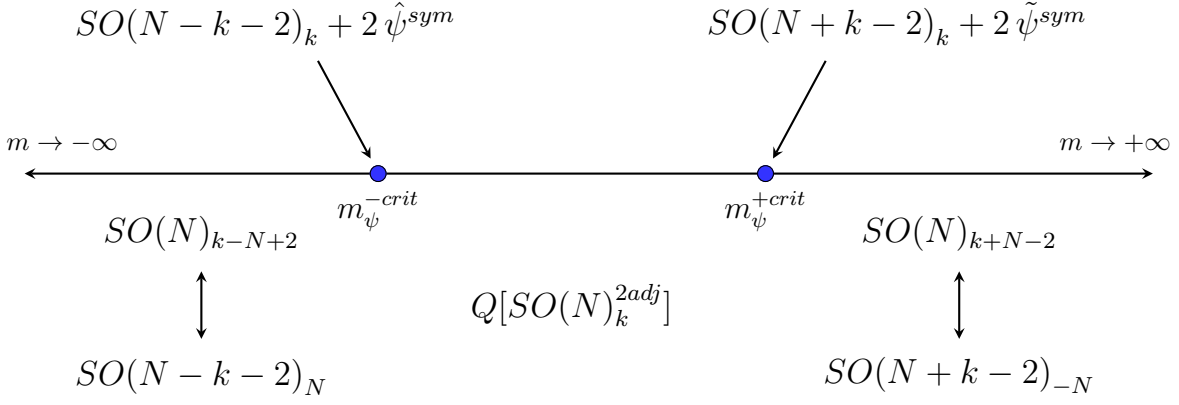

We could construct the phase diagram for the orthogonal group similar to the unitary case. Here, we only illustrate the case when gauge group is $SO(N)$, while generalization for the various covering group $O(N)^p_{K,L}$ could be done similarly as in \cite{Cordova:2017vab}. Construction is similar to the single adjoint case, where the main difference with unitary case is that the representation of the matter is transposed along the duality in accord with single adjoint case in \cite{Gomis:2017ixy}. Hence we propose the following two phase diagram for the $SO(N)$ gauge group with two adjoint fermions as in the figure \ref{fig:so(n)adjsmallk}.

\medskip
\noindent
$\bullet$ {\it $N/2\leq k \leq N-2$}
\medskip

The phase diagram of two adjoint fermions under orthogonal gauge group in the quantum phase regime is presented in figure \ref{fig:so(n)adjsmallk}. Again, there are two distinct phase transitions at $m_\psi = m_\psi^{\pm crit}$, where dual description at $m_\psi^{+crit}$ is always strongly coupled while $m_\psi^{+crit}$ is weakly coupled when $N/2\leq k \leq N-2$. In this case, we can directly obtain the intermediate quantum phase $Q[SO(N)_k^{2adj}]$ from the left dual description :

\begin{equation} \label{eq:so(n)adjqphasesimple}
    \begin{aligned}
    Q[SO(N)_k^{2adj}]=SO(N-k-2)_{2k-N},~ N/2 \leq k <N-2.
    \end{aligned}
\end{equation}

\medskip
\noindent
$\bullet$ {\it $0<k<N/2$ : Duality Chain}
\medskip

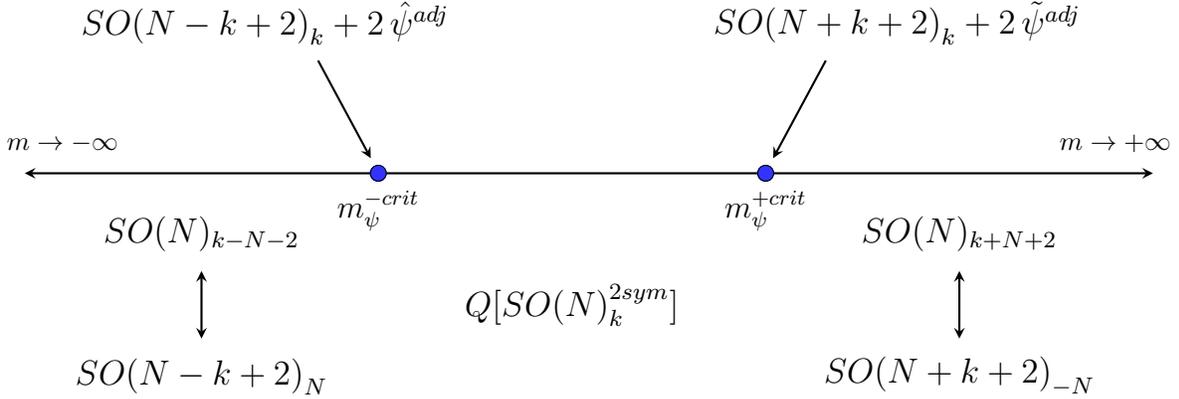
\begin{figure}[!h]
\centering
\begin{tikzpicture}
\node[scale=1.1] at (5.5,4) {$SO(N)_k+2\,\psi^{sym}$};
\node at (13,4) {$0<k<N+2$};
\draw (4.1,3.65) -- (6.9,3.65);
\draw[thick,<->,>=stealth] (-1.5,0) -- (13.5,0);
\filldraw[white!20!blue] (3.2,0) circle (3pt);\draw (3.2,0) circle (3pt);
\filldraw[white!20!blue] (8+.35,0) circle (3pt);\draw (8+.35,0) circle (3pt);
\node at (3.2,-0.5) {$ m_\psi^{-crit}$};
\node at (8+.35,-0.5) {$ m_\psi^{+crit}$};

\node[scale=1.1] at (1.7,2) {$ SO\!\left(N-k+2\right)_{k}+2\,\hat\psi^{adj}$};
\node[scale=1.1] at (10.1,2) {$SO\!\left(N+k+2\right)_{k}+2\,\tilde\psi^{adj}$};
\draw[thick,->,>=stealth] (2.4,1.5) -- (3.1,.2);
\draw[thick,->,>=stealth] (8.8+.35,1.5) -- (8.1+.35,.2);

\node[scale=1.1] at (.85,-.8) {$SO(N)_{k-N-2}$};
\draw[thick,<->,>=stealth] (.85,-1.3) -- (.85,-2.2);
\node[scale=1.1] at (.85,-2.7) {$SO\!\left(N-k+2\right)_{N}$};

\node[scale=1.1] at (5.7+.075,-1.75) {$Q[SO(N)_k^{2sym}]$};
\node[scale=1.1] at (10.925,-.8) {$SO(N)_{k+N+2}$};
\draw[thick,<->,>=stealth] (10.925,-1.3) -- (10.925,-2.2);
\node[scale=1.1] at (10.925,-2.7) {$SO\!\left(N+k+2\right)_{-N}$};

\node at (13,.4) {\footnotesize$m\to+\infty$};
\node at (-1,.4) {\footnotesize$m\to-\infty$};
\end{tikzpicture}
\caption{Phase diagram of $SO(N)$ with two symmetric-traceless fermions for $0<k<N+2$. The right dual description is always strongly coupled and left dual description is weakly coupled only when $N/2 \leq k <N+2$. }
\label{fig:so(n)symsmallk}
\end{figure}

When Chern-Simons level is sufficiently small($0<k<N/2$), the left dual description is also strongly coupled. Similar to the unitary case, we could use the concept of duality chain to determine the quantum phase $Q[SO(N)_k^{2adj}]$. But since the dual fermions belong to the different representation, it is necessary to propose the similar quantum phase diagram for the two symmetric-traceless fermions in the orthogonal gauge group which presented in the figure \ref{fig:so(n)symsmallk}.

Thus pattern of duality chain for the orthogonal group is alternate, where the matter in the adjoint and symmetric-traceless representations are exchanged under each step of duality chain which cover the same intermediate quantum phase $SO(N)_k^{2adj}$. Under this algorithm, the general expression for the quantum phase of $SO(N)_k+2~\psi^{adj}$ for $0<k<N-2$ could be obtained as follows, where the $n^*$ is the number of steps of duality chain we had to apply to identify the quantum phase :

\beq \label{eq:qphaseSO2adj}
~&Q[SO(N)_k^{2adj}]=SO\left(N-n^*k-(1-(-1)^{n^*})\right)_{-N+(n^*+1)k+(1-(-1)^{n^*+1})}
\\&\text{with } n^*\equiv \text{min}\{n=\mathbb Z \cup \{0\} ~\vert~ nk+(1-(-1)^{n}) < N\leq (n+1)k+(1-(-1)^{n+1}) \}
\eeq

Since fermions under adjoint and symmetric representations appear alternately in the duality chain, we could also specify the quantum phase for the $SO(N)_k+2~\psi^{sym}$ similarly as follows with the new definition of $n^*$ corresponding to the number of steps of duality chain in this case\footnote{ \label{fn1} It is necessary to comment that only for the special case of $SO(2)_2+2~\psi^{sym}$ the dual description obtained from a naive application of the duality chain $SO(2)_2+2~\psi^{asym}$ doesn't preserve global symmetry, namely $\mathbb Z_2^{\mathcal C}\times \mathbb Z_2^{\mathcal M}$ symmetry. The main reason is the lack of gauge invariant monopole operator in the dual side due to the decoupled fermions and non-zero Chern-Simons level. While naive continuation of the duality chain predicts $U(1)_2$, in the \ref{sec:specialcase} we propose a reasonable candidate of the quantum phase.}: 

\beq \label{eq:qphaseSO2sym}
~&Q[SO(N)_k^{2sym}]=SO\left (N-n^*k+(1-(-1)^{n^*})\right )_{-N+(n^*+1)k-(1-(-1)^{n^*+1})}
\\&\text{with } n^*\equiv \text{min}\{n=\mathbb Z \cup \{0\} ~\vert~ nk-(1-(-1)^{n}) < N\leq (n+1)k-(1-(-1)^{n+1}) \}
\eeq

\subsection{Quantum Phase for $G=Sp(N)$}

\begin{figure}[!h]
\centering
\begin{tikzpicture}
\node[scale=1.1] at (5.5,4) {$Sp(N)_k+2\,\psi^{adj}$};
\node at (13,4) {$0<k<N+1$};
\draw (4.1,3.65) -- (6.9,3.65);
\draw[thick,<->,>=stealth] (-1.5,0) -- (13.5,0);
\filldraw[white!20!blue] (3.2,0) circle (3pt);\draw (3.2,0) circle (3pt);
\filldraw[white!20!blue] (8+.35,0) circle (3pt);\draw (8+.35,0) circle (3pt);
\node at (3.2,-0.5) {$ m_\psi^{-crit}$};
\node at (8+.35,-0.5) {$ m_\psi^{+crit}$};

\node[scale=1.1] at (1.7,2) {$ Sp\!\left(N-k+1\right)_{k}+2\,\hat\psi^{asym}$};
\node[scale=1.1] at (10.1,2) {$Sp\!\left(N+k+1\right)_{k}+2\,\tilde\psi^{asym}$};
\draw[thick,->,>=stealth] (2.4,1.5) -- (3.1,.2);
\draw[thick,->,>=stealth] (8.8+.35,1.5) -- (8.1+.35,.2);

\node[scale=1.1] at (.85,-.8) {$Sp(N)_{k-N-1}$};
\draw[thick,<->,>=stealth] (.85,-1.3) -- (.85,-2.2);
\node[scale=1.1] at (.85,-2.7) {$Sp\!\left(N-k+1\right)_{N}$};

\node[scale=1.1] at (5.7+.075,-1.75) {$Q[Sp(N)_k^{2adj}]$};
\node[scale=1.1] at (10.925,-.8) {$Sp(N)_{k+N+1}$};
\draw[thick,<->,>=stealth] (10.925,-1.3) -- (10.925,-2.2);
\node[scale=1.1] at (10.925,-2.7) {$Sp\!\left(N+k+1\right)_{-N}$};

\node at (13,.4) {\footnotesize$m\to+\infty$};
\node at (-1,.4) {\footnotesize$m\to-\infty$};
\end{tikzpicture}
\caption{Phase diagram of $Sp(N)$ with two real adjoint fermions for $0<k<N+1$. The right dual description is always strongly coupled and left dual description is weakly coupled only when $N/2 \leq k<N+1$. }
\label{fig:sp(n)adjsmallk}
\end{figure}
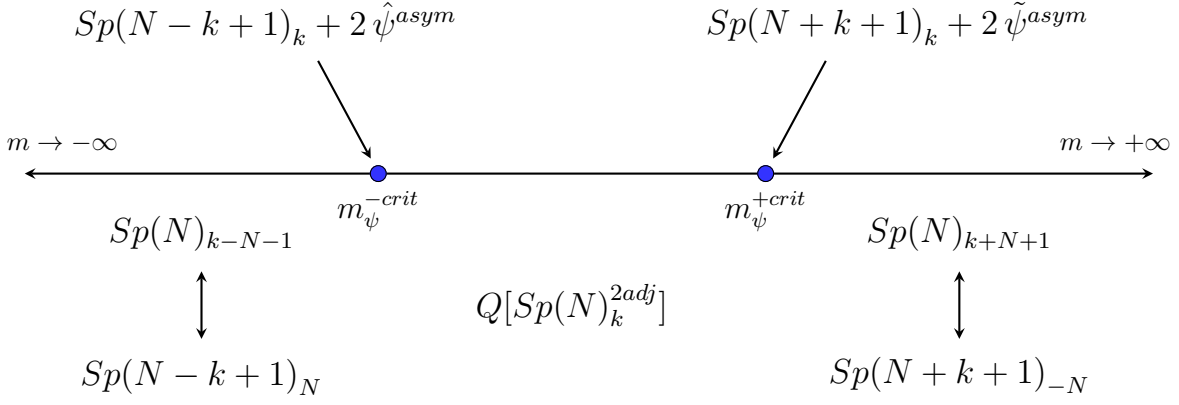

\begin{figure}[!h]
\centering
\begin{tikzpicture}
\node[scale=1.1] at (5.5,4) {$Sp(N)_k+2\,\psi^{asym}$};
\node at (13,4) {$0<k<N-1$};
\draw (4.1,3.65) -- (6.9,3.65);
\draw[thick,<->,>=stealth] (-1.5,0) -- (13.5,0);
\filldraw[white!20!blue] (3.2,0) circle (3pt);\draw (3.2,0) circle (3pt);
\filldraw[white!20!blue] (8+.35,0) circle (3pt);\draw (8+.35,0) circle (3pt);
\node at (3.2,-0.5) {$ m_\psi^{-crit}$};
\node at (8+.35,-0.5) {$ m_\psi^{+crit}$};

\node[scale=1.1] at (1.7,2) {$ Sp\!\left(N-k-1\right)_{k}+2\,\hat\psi^{adj}$};
\node[scale=1.1] at (10.1,2) {$Sp\!\left(N+k-1\right)_{k}+2\,\tilde\psi^{adj}$};
\draw[thick,->,>=stealth] (2.4,1.5) -- (3.1,.2);
\draw[thick,->,>=stealth] (8.8+.35,1.5) -- (8.1+.35,.2);

\node[scale=1.1] at (.85,-.8) {$Sp(N)_{k-N+1}$};
\draw[thick,<->,>=stealth] (.85,-1.3) -- (.85,-2.2);
\node[scale=1.1] at (.85,-2.7) {$Sp\!\left(N-k-1\right)_{N}$};

\node[scale=1.1] at (5.7+.075,-1.75) {$Q[Sp(N)_k^{2asym}]$};
\node[scale=1.1] at (10.925,-.8) {$Sp(N)_{k+N-1}$};
\draw[thick,<->,>=stealth] (10.925,-1.3) -- (10.925,-2.2);
\node[scale=1.1] at (10.925,-2.7) {$Sp\!\left(N+k-1\right)_{-N}$};

\node at (13,.4) {\footnotesize$m\to+\infty$};
\node at (-1,.4) {\footnotesize$m\to-\infty$};
\end{tikzpicture}
\caption{Phase diagram of $Sp(N)$ with two antisymmetric-traceless fermions for $0<k<N-1$. The right dual description is always strongly coupled and left dual description is weakly coupled only when $N/2 \leq k <N-1$. }
\label{fig:sp(n)asymsmallk}
\end{figure}
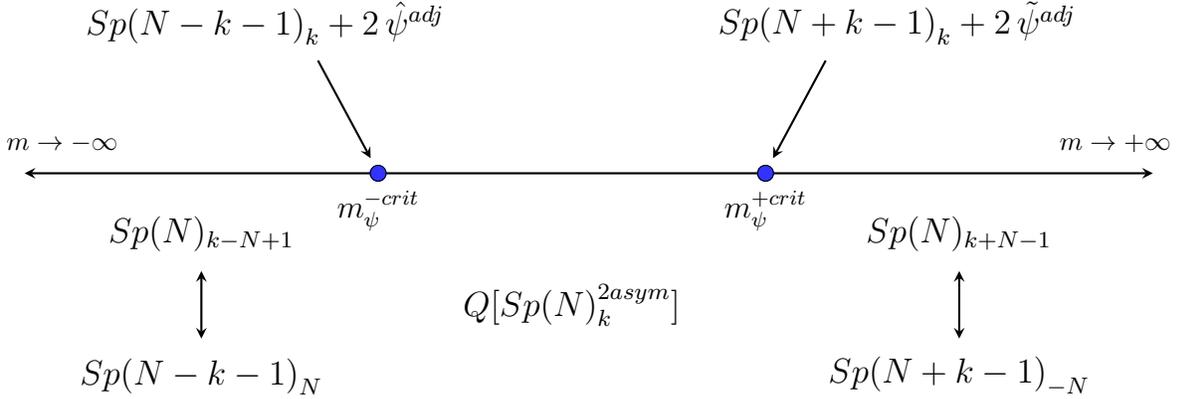

When the gauge group is symplectic, parallel analysis as orthogonal group can be done where we have to alternate adjoint and antisymmetric-traceless representation in this case. Now the quantum phase diagram with two adjoints in figure \ref{fig:sp(n)adjsmallk} and two antisymmetric fermions in figure \ref{fig:sp(n)asymsmallk} would complete the duality chain.

Parallel to the orthogonal group case, quantum phase can be determined as follows under the $n^*$ steps of duality chain :

\beq \label{eq:qphaseSp2adj}
~&Q[Sp(N)_k^{2adj}]=Sp\left (N-n^*k+\frac{1}{2}(1-(-1)^{n^*})\right )_{-N+(n^*+1)k-\frac{1}{2}(1-(-1)^{n^*+1})}
\\&\text{with } n^*\equiv \text{min}\{n=\mathbb Z \cup \{0\} ~\vert~ nk-\frac{1}{2}(1-(-1)^{n}) < N\leq (n+1)k-\frac{1}{2}(1-(-1)^{n+1}) \}
\eeq

\beq \label{eq:qphaseSp2asym}
~&Q[Sp(N)_k^{2asym}]=Sp\left(N-n^*k-\frac{1}{2}(1-(-1)^{n^*})\right)_{-N+(n^*+1)k+\frac{1}{2}(1-(-1)^{n^*+1})}
\\&\text{with } n^*\equiv \text{min}\{n=\mathbb Z \cup \{0\} ~\vert~ nk+\frac{1}{2}(1-(-1)^{n}) < N\leq (n+1)k+\frac{1}{2}(1-(-1)^{n+1}) \}
\eeq

\subsection{Phase with Spontaneously Broken Partial 1-form, 0-form Symmetry}

\medskip
\noindent
$\bullet$ {\it Phase with partial spontaneous breaking of $\mathbb Z_N$ 1-form symmetry in $SU(N)$ gauge theory}
\medskip

We could see the interesting dynmical implication of the $SU(N)_k+2~\psi^{adj}$ theory when we look at the fate of 1-form symmetry in the quantum phase of $Q[SU(N)_k^{2adj}]$ given by \eqref{eq:su(n)qphase} :

\beq
Q[SU(N)_k^{2adj}]=(ST^2)^{\ceil{\frac{N}{k}}-2}U\!\left(N-\ceil{\frac{N}{k}}k+k\right)_{-N+\ceil{\frac{N}{k}}k,N-\ceil{\frac{N}{k}}k+2k}
\eeq

We see that when $k \nmid N$, the TQFT $Q[SU(N)_k^{2adj}]$ has $\mathbb Z_{N}$ 1-form global symmetry. But when $k\mid N$, we see that 1-form symmetry is reduces to $\mathbb Z_{N} /\mathbb Z_{k}$ in the IR due to the confinement of non-abelian Yang-Mills theory $SU(k)_0$ for $k\geq 0$. Thus we see that when $k\mid N$, only the nontrivial elements of $\mathbb Z_{N}/\mathbb Z_{k}$ 1-form symmetry are spontaneously broken in the IR while the $\mathbb Z_{k}\subset \mathbb Z_N$ is confined. This has the following implication that when we gauge the anomaly free subgroup of the $\mathbb Z_{k}\subset \mathbb Z_N$ 1-form symmetry in the $SU(N)_k+2 ~\psi^{adj}$ for $k \mid N$, then the theory flows to a phase with spontaneously broken discrete vacua coupled to TQFT.

\medskip
\noindent
$\bullet$ {\it Phase with spontaneously broken $\mathbb Z_2^{\mathcal C}$, $\mathbb Z_2^{\mathcal M}$ in the $SO(N)$ gauge theory}
\medskip

The duality chain combined with the fact that $SO(N)_0$ with $N\geq 3$ (or $SO(2)_0$ with double monopole deformation.) exhibit the spontaneous breaking of $\mathbb Z_2^{\mathcal M}$ magnetic symmetry has following implication. Since the charge conjugation symmetry $\mathcal C$ and magnetic symmetry $\mathcal M$ is interchanged under the each step of duality chain similar to \cite{Cordova:2017vab}, we have following special phases for the $SO(N)$ gauge theory :

\begin{equation}
\begin{gathered}
 SO\left (2mn\right)_{n}+2~\psi^{sym} ~~~\&~~~ SO\left (2mn+6\right )_{n+3}+2~\psi^{adj}, ~~m,n\in\mathbb N 
\\ \longrightarrow \text{Two trivial vacua from a spontaneously breaking of }\mathbb Z_2^{\mathcal C}
\end{gathered}
\end{equation}

\begin{equation}
\begin{gathered}
SO\left (2mn+6m+n+1\right )_n+2~\psi^{sym} ~~~\&~~~ SO\left(2mn+n+2\right)_n+2~\psi^{adj} , ~~m,n\in \mathbb N 
\\ \longrightarrow \text{Two trivial vacua from a spontaneous breaking of }\mathbb Z_2^{\mathcal M}
\end{gathered}
\end{equation}

\section{More Duality Chains and Quantum Phases} \label{sec:more dualities}

\subsection{$G_k$+Pair of Rank-Two/Adjoint Fermions}

Here, we list the possible generalizations of the quantum phase with duality chain for any two combination of rank-two/adjoint fermions under the gauge group $SU(N),SO(N),Sp(N)$ with positive $k$ without loss of generality. In parallel to the two adjoints case, it is natural to conjecture that quantum phase exist only when $k<T(R)$, where $T(R)$ is Dynkin index of total representation of the matter which is the direct sum of the each fermion's representation.

\medskip
\noindent
$\bullet$ {\it $U(N)_{k,k}+~2~\psi^{adj}$}
\medskip

We first point out that there is a great simplification when 2 adjoints fermions are charged under $U(N)$ gauge group, where generalized level/rank duality is no more required. We explain the connection between the case of $SU(N)$ gauge group through $SL(2,\mathbb Z)$ transformation in section \ref{subsec:sl(2,z)selfcons}. Duality chain exists for $0<k<N$ as $SU(N)$ case, with the following quantum phase :

\smallskip

\begin{equation}
\begin{gathered}
U\left(N\right)_{k,k}+ ~2~\psi^{adj},~ m_\psi=m^{+crit}_\psi \longleftrightarrow~~ U\!\left(N+k\right)_{k,k}+~2~\tilde \psi^{adj},~ m_{\tilde \psi}=m^{-crit}_{\tilde \psi} \\[+5pt]
U\left (N\right)_{k,k}+~2~\psi^{adj},~ m_{\psi}=m^{-crit}_{\psi} \longleftrightarrow~~ U\!\left(N-k\right)_{k,k}+~2~\hat \psi^{adj},~ m_{\hat \psi}=m^{+crit}_{\hat \psi}\,.
\\ Q[U(N)_{k,k}^{2adj}]=U\!\left(N-\ceil{\frac{N}{k}}k+k\right)_{-N+\ceil{\frac{N}{k}}k,k} 
\end{gathered}
\end{equation}

\medskip
\noindent
$\bullet$ {\it $SU(N)_k+\psi^{sym}+\psi^{asym}$}
\medskip

This is the case when there are one symmetric and one antisymmetric representation of fermion. Duality chain exists for $0<k<N$, with the following quantum phase\footnote{Here the how does $\psi^{sym/asym}$ is charged under multiple $U(1)$ gauge fields can be established using the consistency under the $SL(2,\mathbb Z)$ operation similar to the discussion in \ref{subsec:sl(2,z)selfcons}. Straightforward analysis shows that $\psi^{sym/asym}$ is neutral under any additional $U(1)$ gauge field generated from $SL(2,\mathbb Z)$ operations.}

\smallskip

\begin{equation}
\begin{gathered}
SU(N)_k+ \psi^{sym}+\psi^{asym},~ m_\psi=m^{+crit}_\psi \longleftrightarrow~~ U\!\left(N+k\right)_{k,2k+N}+\tilde \psi^{sym}+\tilde \psi^{asym},~ m_{\tilde \psi}=m^{-crit}_{\tilde \psi} \\[+5pt]
SU(N)_k+\psi^{sym}+\psi^{asym},~ m_{\psi}=m^{-crit}_{\psi} \longleftrightarrow~~ U\!\left(N-k\right)_{k,2k-N}+\hat \psi^{sym}+\hat \psi^{asym},~ m_{\hat \psi}=m^{+crit}_{\hat \psi}\,.
\\  Q[SU(N)_k^{sym+asym}]= (ST^{-2})^{\ceil{\frac{N}{k}}-2}U\!\left(N-\ceil{\frac{N}{k}}k+k\right)_{-N+\ceil{\frac{N}{k}}k,-3N+3\ceil{\frac{N}{k}}k-2k}
\end{gathered}
\end{equation}

\medskip
\noindent
$\bullet$ {\it $SU(N)_k+2~\psi^{sym}$ $\&$ $SU(N)_k+2~\psi^{asym}$}
\medskip

Duality chain exists for $0<k<N\pm 2$, for the two symmetric or antisymmetric cases, with the following quantum phase determined under the $n^*_{S/A}$ steps of duality chain :

\begin{equation}
\begin{gathered}
SU(N)_k+ ~2~\psi^{sym/asym},~ m_\psi=m^{+crit}_\psi \longleftrightarrow~~ U\!\left(N+k\pm 2\right)_{k,2k+N\pm 2}+~2~\tilde \psi^{asym/sym},~ m_{\tilde \psi}=m^{-crit}_{\tilde \psi} \\[+5pt]
SU(N)_k+~2~\psi^{sym/asym},~ m_{\psi}=m^{-crit}_{\psi} \longleftrightarrow~~ U\!\left(N-k\pm 2\right)_{k,2k-N\mp2}+~2~\hat \psi^{asym/sym},~ m_{\hat \psi}=m^{+crit}_{\hat \psi}\,.
\\ Q[SU(N)_k^{2sym/2asym}] =
\\(ST^{-2})^{n^*-1}U\!\left(N-n^*_{S/A}k\pm(1-(-1)^{n_{S/A}^*}) \right)_{-N+(n^*_{S/A}+1)k\mp(1-(-1)^{n_{S/A}^*+1})
,-3N+(3n^*_{S/A}+1)k\mp(3-(-1)^{n_{S/A}^*})}
\\ \text{with } n^*_{S/A}\equiv \text{min}\{n=\mathbb Z \cup \{0\} ~\vert~ nk\mp (1-(-1)^{n}) < N\leq (n+1)k\mp (1-(-1)^{n+1}) \}
\end{gathered}
\end{equation}

\medskip
\noindent
$\bullet$ {\it $SU(N)_k+\psi^{sym}+\psi^{adj}$ $\&$ $SU(N)+\psi^{asym}+\psi^{adj}$}
\medskip

Duality chain exists for $0<k<N\pm 1$, for the adjoint+symmetric or adjoint+antisymmetric cases, with the following quantum phase determined under the $n^*_{S/A}$ steps of duality chain :

\begin{equation}
\begin{gathered}
SU(N)_k+\psi^{adj}+ \psi^{sym/asym},~ m_\psi=m^{+crit}_\psi \longleftrightarrow~~ U\!\left(N+k\pm 1 \right)_{k,k}+\tilde \psi^{adj}+\tilde \psi^{asym/sym},~ m_{\tilde \psi}=m^{-crit}_{\tilde \psi} \\[+5pt]
SU(N)_k+\psi^{adj}+\psi^{sym/asym},~ m_{\psi}=m^{-crit}_{\psi} \longleftrightarrow~~ U\!\left(N-k\pm 1\right)_{k,k}+\hat \psi^{adj}+\hat \psi^{asym/sym},~ m_{\hat \psi}=m^{+crit}_{\hat \psi}\,.
\\  Q[SU(N)_k^{adj+sym/adj+asym}] =
\\ (S)^{n^*_{S/A}}SU\!\left(N-n^*_{S/A}k\pm \frac{1}{2}(1-(-1)^{n_{S/A}^*}) \right)_{-N+(n^*_{S/A}+1)k\mp\frac{1}{2}(1-(-1)^{n_{S/A}^*+1})}
\\ \text{with } n^*_{S/A}\equiv \text{min}\{n=\mathbb Z \cup \{0\} ~\vert~ nk\mp \frac{1}{2}(1-(-1)^{n}) < N\leq (n+1)k\mp \frac{1}{2}(1-(-1)^{n+1}) \}
\end{gathered}
\end{equation}

Where $S$ is the $SL(2,\mathbb Z)$ operation defined in the appendix \ref{appsec:genlevelrank}.

\medskip
\noindent
$\bullet$ {\it $SO(N)_k+\psi^{adj}+\psi^{sym}$}
\medskip

Duality chain exists for $0<k<N$, with the following quantum phase :

\begin{equation}
\begin{gathered}
SO(N)_k+\psi^{adj}+ \psi^{sym},~ m_\psi=m^{+crit}_\psi \longleftrightarrow~~ SO(N+k)_k+\tilde \psi^{adj}+\tilde \psi^{sym},~ m_{\tilde \psi}=m^{-crit}_{\tilde \psi} \\[+5pt]
SO(N)_k+\psi^{adj}+\psi^{sym},~ m_{\psi}=m^{-crit}_{\psi} \longleftrightarrow~~ SO(N-k)_k+\hat \psi^{adj}+\hat \psi^{sym},~ m_{\hat \psi}=m^{+crit}_{\hat \psi}\,.
\\  Q[SO(N)_k^{adj+sym}] =SO(N-\ceil{\frac{N}{k}}k+k)_{N-\ceil{\frac{N}{k}}k}
\end{gathered}
\end{equation}

\medskip
\noindent
$\bullet$ {\it $Sp(N)_k+\psi^{adj}+\psi^{asym}$}
\medskip

Duality chain exists for $0<k<N$, with the following quantum phase :

\begin{equation}
\begin{gathered}
Sp(N)_k+\psi^{adj}+ \psi^{asym},~ m_\psi=m^{+crit}_\psi \longleftrightarrow~~ Sp(N+k)_k+\tilde \psi^{adj}+\tilde \psi^{asym},~ m_{\tilde \psi}=m^{-crit}_{\tilde \psi} \\[+5pt]
Sp(N)_k+\psi^{adj}+\psi^{asym},~ m_{\psi}=m^{-crit}_{\psi} \longleftrightarrow~~ Sp(N-k)_k+\hat \psi^{adj}+\hat \psi^{asym},~ m_{\hat \psi}=m^{+crit}_{\hat \psi}\,.
\\  Q[Sp(N)_k^{adj+asym}] =Sp(N-\ceil{\frac{N}{k}}k+k)_{N-\ceil{\frac{N}{k}}k}
\end{gathered}
\end{equation}

\subsection{$G_{k_1}\times G_{k_2}$+Two Bifundamental Fermions} \label{subsec:two-bifund}

Surprisingly, duality chain also exists for the 2-node quiver theory with the two bifundamental fermions case. Here we only consider when the both gauge group is type of $SU\times SU,~SO\times SO,~Sp\times Sp$. Story is very similar to the general pair of rank-two/adjoint fermions so far treated above. When there is a single bifundamental fermion, 'weakly coupled' dual descriptions are exist which shares the same intermediate phase by level/rank duality\cite{Aitken:2019shs}. In the presence of two bifundamental fermions, direct description of phase transition in terms of weakly coupled theory is not manageable, and it is also natural to conjecture that similar duality chain would give the information about the quantum phase. The notable speciality of the two bifundamentals theory is that we could match the gravitational counterterms along the any step of duality chain exactly as shown in the section \ref{subsec:gravitational}.

Similarly, it is natural to conjecture that quantum phase exist for the $G_{k_1}\times G_{k_2}+2~\psi^{bifund}$ when the $\vert k_1\vert <h_2, ~\vert k_2\vert <h_1$ where $h_{1,2}$ is the dual coxeter number of first and second gauge group. Thus the first and second gauge group's rank are changed by $k_2,~k_1$ under the each step of duality chain\footnote{Note that for $k_1 k_2 <0$, chain makes one of the rank decrease while other increase which is the unique feature of 2-node quiver theory compared to the simple gauge group case.}. When $k_1=k_2=0$, duality chain does not exist similar to the adjoint/rank-two cases before. The structure of duality chains and quantum phases are following\footnote{For the introduction to the quiver Chern-Simons theories, see for example \cite{Jensen:2017dso}.}(We use $U(N)_k\equiv U(N)_{k,k}$ for simplicity) :

\medskip
\noindent
$\bullet$ {\it $SU(N_1)_{k_1}\times U(N_2)_{k_2}+2~\psi^{bifund}$}
\medskip

Here we choose the $SU\times U$ quiver for simplicity. For the case of $SU\times SU$, $U\times U$ can be obtained from appropriate $SL(2,\mathbb Z)$ operation on the $SU\times U$ results. The dualities and quantum phase are given as following :

\beq
SU(N_1)_{k_1}\times U(N_2)_{k_2}&+2~\psi^{bf}, ~ m_\psi=m^{+crit}_\psi
\\&\longleftrightarrow  SU(N_1+k_2)_{k_1}\times U(N_2+k_1)_{k_2} +2~\tilde \psi^{bf},~ m_{\hat \psi}=m^{-crit}_{\hat \psi}
\\SU(N_1)_{k_1}\times U(N_2)_{k_2}&+2~\psi^{bf}, ~ m_\psi=m^{-crit}_\psi
\\&\longleftrightarrow  SU(N_1-k_2)_{k_1}\times U(N_2-k_1)_{k_2} +2~{\hat \psi}^{bf},~ m_{{\hat \psi}}=m^{+crit}_{{\hat \psi}}
\eeq

\begin{equation}
\begin{gathered}
Q[(SU(N_1)_{k_1}\times U(N_2)_{k_2})^{2bifund}]=
\\
\begin{cases}
SU\left (N_1-\ceil{\frac{N_1}{k_2}}k_2+k_2\right)_{-N_2+\ceil{\frac{N_1}{k_2}}k_1}\times U\left (N_2-\ceil{\frac{N_1}{k_2}}k_1+k_1\right )_{-N_1+\ceil{\frac{N_1}{k_2}}k_2}
&:~\ceil{\frac{N_1}{k_2}}\leq \ceil{\frac{N_2}{k_1}}
\\SU\left (N_1-\ceil{\frac{N_2}{k_1}}k_2+k_2\right)_{-N_2+\ceil{\frac{N_2}{k_1}}k_1}\times U\left (N_2-\ceil{\frac{N_2}{k_1}}k_1+k_1\right )_{-N_1+\ceil{\frac{N_2}{k_1}}k_2}
&:~\ceil{\frac{N_2}{k_1}}\leq \ceil{\frac{N_1}{k_2}}
\end{cases}
\end{gathered}
\end{equation}

\medskip
\noindent
$\bullet$ {\it $SO(N_1)_{k_1}\times SO(N_2)_{k_2}+2~\psi^{bifund}$}
\medskip

One can do the similar analysis for the SO group where now the situation is more simpler because of the absence of $U(1)$ factors compared to the unitary group case. Duality chain and quantum phase are following :

\beq
SO(N_1)_{k_1}\times SO(N_2)_{k_2}&+2~\psi^{bf}, ~ m_\psi=m^{+crit}_\psi
\\&\longleftrightarrow  SO(N_1+k_2)_{k_1}\times SO(N_2+k_1)_{k_2} +2~\tilde \psi^{bf},~ m_{\hat \psi}=m^{-crit}_{\hat \psi}
\\SO(N_1)_{k_1}\times SO(N_2)_{k_2}&+2~\psi^{bf}, ~ m_\psi=m^{-crit}_\psi
\\&\longleftrightarrow  SO(N_1-k_2)_{k_1}\times SO(N_2-k_1)_{k_2} +2~{\hat \psi}^{bf},~ m_{{\hat \psi}}=m^{+crit}_{{\hat \psi}}
\eeq

\begin{equation}
\begin{gathered}
Q[(SO(N)_{k_1}\times SO(N_2)_{k_2})^{2bifund}]=
\\
\begin{cases}
SO\left (N_1-\ceil{\frac{N_1}{k_2}}k_2+k_2\right)_{-N_2+\ceil{\frac{N_1}{k_2}}k_1}\times SO\left (N_2-\ceil{\frac{N_1}{k_2}}k_1+k_1\right )_{-N_1+\ceil{\frac{N_1}{k_2}}k_2}
& :~\ceil{\frac{N_1}{k_2}}\leq \ceil{\frac{N_2}{k_1}}
\\SO\left (N_1-\ceil{\frac{N_2}{k_1}}k_2+k_2\right)_{-N_2+\ceil{\frac{N_2}{k_1}}k_1}\times SO\left (N_2-\ceil{\frac{N_2}{k_1}}k_1+k_1\right )_{-N_1+\ceil{\frac{N_2}{k_1}}k_2}
& :~\ceil{\frac{N_2}{k_1}}\leq \ceil{\frac{N_1}{k_2}}
\end{cases}
\end{gathered}
\end{equation}

\medskip
\noindent
$\bullet$ {\it $Sp(N_1)_{k_1}\times Sp(N_2)_{k_2}+2~\psi^{bifund}$}
\medskip

One can do the similar analysis for the symplectic group. Duality chain and quantum phase is following :

\beq
Sp(N_1)_{k_1}\times Sp(N_2)_{k_2}&+2~\psi^{bf}, ~ m_\psi=m^{+crit}_\psi
\\&\longleftrightarrow  Sp(N_1+k_2)_{k_1}\times Sp(N_2+k_1)_{k_2} +2~\tilde \psi^{bf},~ m_{\hat \psi}=m^{-crit}_{\hat \psi}
\\Sp(N_1)_{k_1}\times Sp(N_2)_{k_2}&+2~\psi^{bf}, ~ m_\psi=m^{-crit}_\psi
\\&\longleftrightarrow  Sp(N_1-k_2)_{k_1}\times Sp(N_2-k_1)_{k_2} +2~{\hat \psi}^{bf},~ m_{{\hat \psi}}=m^{+crit}_{{\hat \psi}}
\eeq

\begin{equation}
\begin{gathered}
Q[(Sp(N)_{k_1}\times Sp(N_2)_{k_2})^{2bifund}]=
\\
\begin{cases}
Sp\left (N_1-\ceil{\frac{N_1}{k_2}}k_2+k_2\right)_{-N_2+\ceil{\frac{N_1}{k_2}}k_1}\times Sp\left (N_2-\ceil{\frac{N_1}{k_2}}k_1+k_1\right )_{-N_1+\ceil{\frac{N_1}{k_2}}k_2}
& :~\ceil{\frac{N_1}{k_2}}\leq \ceil{\frac{N_2}{k_1}}
\\Sp\left (N_1-\ceil{\frac{N_2}{k_1}}k_2+k_2\right)_{-N_2+\ceil{\frac{N_2}{k_1}}k_1}\times Sp\left (N_2-\ceil{\frac{N_2}{k_1}}k_1+k_1\right )_{-N_1+\ceil{\frac{N_2}{k_1}}k_2}
& :~\ceil{\frac{N_2}{k_1}}\leq \ceil{\frac{N_1}{k_2}}
\end{cases}
\end{gathered}
\end{equation}

\section{Consistency Checks} \label{sec:additional_consistency_checks}

\subsection{Deformations under RG flows}

Here we break the $U(1)_B$ flavor symmetry in the two adjoints duality \eqref{eq:su(n)smallk} by giving mass deformation to the single flavor and discover that renormalization group flows of the duality \eqref{eq:su(n)smallk} is consistent with the duality of the single real adjoint fermion analyzed in \cite{Gomis:2017ixy}. We only illustrate the $SU$ gauge group while the $SO/Sp$ cases together with the generalizations in the section \ref{sec:more dualities} could be done similarly.

We first discuss what kind of mass deformations in the phase diagram of figure \ref{fig:su(n)smallk} are manageable semi-classically. Recall that the original theory has fermions $\psi^{adj}_{1,2}$, and left/right dual description has fermions ${\hat\psi}^{adj}_{1,2}/{\tilde \psi}^{adj}_{1,2}$ and mass deformations are related by $\delta m_\psi=-\delta m_{\hat\psi}$ and $\delta m_\psi=-\delta m_{\tilde\psi}$. Since two asymptotic phases at $m_\psi=\pm \infty$ is described by semiclassical analaysis while the intermediate quantum phase is governed by strongly coupled interactions, the only deformation that we could access semiclassically are $\delta m_{\psi_1}=-\delta m_{{\hat \psi}_1} \ll - \Lambda_{QCD}$ along the left critical point $m_{\psi}=m_\psi^{-crit}$ and independently $\delta m_{\psi_1}=-\delta m_{{\tilde \psi}_1} \gg \Lambda_{QCD}$ along the right critical point $m_{\psi}=m_\psi^{+crit}$. After this independent mass deformations for each duality in \eqref{eq:su(n)smallk}, we get (omitting subscript `2' in the fermion) :

\beq \label{eq:su(n)rgflow}
&SU(N)_{k+N/2}~+~ \psi^{adj} &\longleftrightarrow~~ &U\!\left(N+k\right)_{-N/2+k/2,-N}~+~ \tilde \psi^{adj} \\[+5pt]
&SU(N)_{k-N/2}~+~ \psi^{adj}&\longleftrightarrow~~ &U\!\left(N-k\right)_{N/2+k/2,N}~+~\hat \psi^{adj}.
\eeq

Since the range of the original duality in \eqref{eq:su(n)smallk} is $0<k<N$, we could see that above \eqref{eq:su(n)rgflow} is nothing but the duality between single adjoint fermion discussed in \cite{Gomis:2017ixy}. It's become transparent when we change the variables to the each equations in \eqref{eq:su(n)rgflow} respectively :

\beq \label{eq:su(n)1adj}
SU(\tilde N/2+\tilde k)_{3\tilde N/4-\tilde k/2}~+~ \psi^{adj}~ &\longleftrightarrow~~ U(\tilde N)_{-\tilde k,\tilde N/2+\tilde k}~+~ \tilde \psi^{adj} ~  &0<\tilde k< \tilde N/2
\\[+5pt]
SU(N)_{\hat k}~+~ \psi^{adj}~&\longleftrightarrow~~ U(N/2-\hat k)_{\hat k/2+3N/4, N}~+~\hat \psi^{adj},~ &-N/2 <\hat k <N/2.
\eeq

Where the second line is its original form, while the first one is the version where appropriate $SL(2,\mathbb Z)$ operation to the phase diagram of single adjoint fermion is performed. Both together complete the phase diagram of single real adjoint fermions in \cite{Gomis:2017ixy}.

\subsection{Matching $SU(N)_1+2~\psi^{adj}$ Phase from Two Approaches}

The consistency established in this section highly supports the appearance of generalize level-rank duality in \ref{appsec:genlevelrank} for the unitary group case. For the case $SU(N)_1+2~\psi^{adj}$, we could find a consistent weakly-coupled dual description from semi-classical reasoning. On the other hand, duality chain predict the quantum phase to be non-trivial abelian CS theory with K-matrix description. It turns out that both phases are dual to each other as we describe now.

\begin{itemize}
    \item From a 2-dimensional phase diagram.
    
    We could vary the bare mass of each adjoint fermion in the $SU(N)$ gauge group independently and see how does phase diagram looks like. Similar to the analysis of \cite{Argurio:2019tvw,Baumgartner:2019frr}, the information from the phase diagram of single adjoint fermion at the boundary of 2-dimensional phase diagram might help us to analyze the inner region near $m_{\psi}=0$. Notably, for the case of $k=1$, all the critical lines at the boundary $\vert m \vert \rightarrow \infty$ are in the semiclassical regime and there is no quantum phase. Thus we could draw the most natural phase diagram as illustrated in the figure \ref{fig:2dphaseadj}.
    
    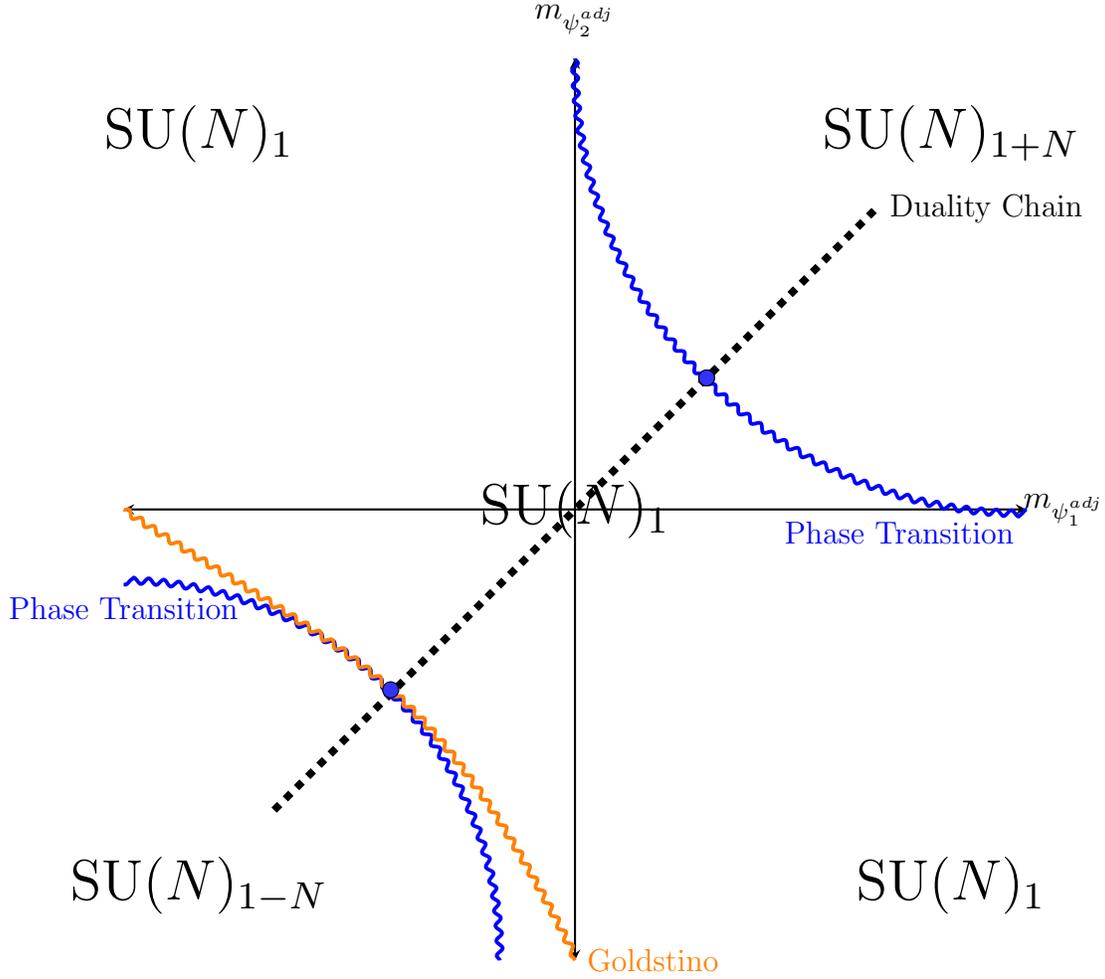
\begin{figure}[!h]
\centering
\begin{tikzpicture}

\draw[thick,<->,>=stealth] (0,0) -- (12,0);
\draw[thick,<->,>=stealth] (6,-6) -- (6,6);


\draw[thick, dashed,line width=1 mm]  (7.5-5.5,1.5-5.5) -- (7.5+2.5,1.5+2.5) node[anchor=west] {Duality Chain};

\draw (5.0,-6.0)[blue,thick,decorate,line width=0.5mm,decoration={snake,amplitude=0.4mm,segment length=2mm}] arc (0:90:5cm)node[anchor=north ] {Phase Transition}; 

\draw [orange,thick,line width=0.5mm,decorate,decoration={snake,amplitude=0.4mm,segment length=2mm}] (0,0) .. controls  (3,-2) and (4,-2) .. (6,-6)  node[anchor=west] {Goldstino};

\draw (6.0,6.0)[ blue,thick,decorate,line width=0.5mm,decoration={snake,amplitude=0.4mm,segment length=2mm}] arc (180:270:6cm)node[anchor=north east] {Phase Transition};

\node[scale=1.8] at (11,5) {$\mathrm{SU}(N)_{1+N}$};
\node[scale=1.8] at (1,-5) {$\mathrm{SU}(N)_{1-N}$};
\node[scale=1.8] at (1,5) {$\mathrm{SU}(N)_{1}$};
\node[scale=1.8] at (11,-5) {$\mathrm{SU}(N)_{1}$};

\node[scale=1.2] at (12.5,0) {\footnotesize$m_{\psi^{adj}_1}$};
\node[scale=1.2] at (6,6.5) {\footnotesize$m_{\psi^{adj}_2}$};

\filldraw[white!20!blue] (3.55,-2.4) circle (3pt);\draw (3.55,-2.4) circle (3pt);

\filldraw[white!20!blue] (7.75,1.75) circle (3pt);\draw (7.75,1.75) circle (3pt);

\node[scale=1.8] at (6,0) {$\mathrm{SU}(N)_{1}$};

\end{tikzpicture}
\caption{Generic shape of the two dimensional phase diagram for $SU(N)_1+2~\psi^{adj}$. The diagonal black line represent the region where duality chain applies. Near the boundary $\vert m_\psi \vert\rightarrow \infty$, we could use the phase diagram of single adjoint fermion reviewed in the section \ref{sec:review} to infer the phase in the middle region. Our case is special since their is no quantum phase at the boundary, thus it is natural to predict that $SU(N)_1$ phase in the 2nd and 4th quadrant are smoothly connected through the middle region. }\label{fig:2dphaseadj}. 
\end{figure} 
    
    Here the most natural candidate for the quantum phase near $m_\psi=0$ is $SU(N)_1=U(1)_{-N}$, mainly because the critical lines of Goldstino could not affect the any IR phase abruptly. Our expectation is independently supported by the consistent mutually non-local weakly coupled dual descriptions along the diagonal mass line (not a duality chain!) as expressed in the table \ref{table:SU(N)_1}. 
    
    \begin{table}[!ht]\label{su2adjlevel1}\begin{tabularx}{\textwidth}{|C|C|C|} \hline
\multicolumn{3}{|c|}{$SU(N)_{1}$+ 2~ $\psi^{adj}$ }\\ 
\hline
$SU(N)_{-N+1}$
&
& $SU(N)_{N+1}$ 
\\
$\updownarrow$ & & \\
$U(N-1)_{N,N}$ 
& $U(N-1)_{1,N}$&
$\updownarrow$ \\
&$\updownarrow$&\\
&$U(N+1)_{1,-N}$&$U(N+1)_{-N,-N}$ \\ 
\hline
\multicolumn{3}{|c|}{~~~~~~~~~~\!~~$\uparrow$ ~~~~~~~~~~~~~~~~~~~~~~~~~~~~~~~~~~~~~$\uparrow$~~~~~~~~~~~~}
\\ 
\multicolumn{3}{|c|}{$U(N-1)_{\frac{N+1}{2},N}+\text{adj} ~\hat \psi$
~~~
$U(N+1)_{-\frac{N-1}{2},-N}+\text{adj} ~\tilde \psi$ 
}
\\
\hline
\end{tabularx}
\caption{phase diagram for $SU(N)_1+2~\psi^{adj}$ }
\label{table:SU(N)_1}
\end{table}

\item From duality chain

Applying the result of the duality chain in \eqref{eq:su(n)qphase} gives the quantum phase $Q[SU(N)_1^{2adj}]$ as a following rank N-1 abelian TQFT :

\beq
&Q[SU(N)_1^{adj}]=(ST^2)^{N-2}U(1)_2=\sum_{i,j=1,\dots,N-1} \frac{k_{ij}}{4\pi}a_ida_j
\\&\text{where } k_{ij}=2\delta_{ij}-\delta_{i,j+1}-\delta_{i,j-1}
\eeq
\end{itemize}

Surprisingly, this abelian TQFT is dual to  $U(1)_{-N}$\cite{frohlich1994integral}.

\subsection{Matching from the Isomorphisms of low-rank Lie groups} \label{sec:specialcase}

Here we uses the various isomorphisms between low rank Lie groups to see whether the quantum phase using the each duality chain gives the consistent result. As we will see, this procedure is highly non-trivial since the each duality chains are totally different in general.\footnote{For the treatment of various discrete gauge fields in the orthogonal gauge groups and its level/rank dualities, see \cite{Cordova:2017vab}.}

\begin{itemize}

    \item  $Spin(4)_1+2~\psi^{adj}=SU(2)_1\times SU(2)_1+2~\psi^{adj}$
    
    We have $Q[Spin(4)_1^{2adj}]=(\mathbb Z_2)_4$, while $Q[SU(2)_1\times SU(2)_1+2 \psi^{adj}]=Q[SU(2)_1^{2adj}]^2=U(1)_{-2}\times U(1)_{-2}$. If we use $(\mathbb Z_2)_4=Spin(4)_{-1}$ we get the same phase.

    \item $Spin(5)_1+2~\psi^{adj}=Sp(2)_1+2~\psi^{adj}$
    
    We have $Q[Spin(5)_1^{2adj}]=O(3)^0_{0,2}= \frac{O(3)^0_{0,0}\times Spin(2)_{-1}}{\mathbb Z_2}=U(1)_1$, while $Q[Sp(2)_1^{adj}]=Sp(2)_0$. So both approaches yields the same trivial phase.    
    \item  $Spin(5)_2+2~\psi^{adj}=Sp(2)_2+2~\psi^{adj}$
    
    We have $Q[Spin(5)_2^{2adj}]=(\mathbb Z_2)_5$, while $Q[Sp(2)_2^{adj}]=Sp(1)_2$. Since $(\mathbb Z_2)_5=spin(3)_1=SU(2)_2$ we get the consistent result.
    \item $SU(4)_3+2~\psi^{adj}=Spin(6)_3+2~\psi^{adj}$
    
    We have $Q[SU(4)_3^{2adj}]=U(1)_4$ while $Q[Spin(6)_3^{adj}]=(\mathbb Z_2)_6=Spin(2)_1$ which indeed gives the same phase.

        \item  $SU(4)_1+2~\psi^{adj}=Spin(6)_1+2~\psi^{adj}$
    
    We have $Q[SU(4)_1^{2adj}]=U(1)_{-4}$ while $Q[Spin(6)_1^{adj}]=(\mathbb Z_2)_2$. If we use $(\mathbb Z_2)_2=U(1)_{-4}$, we see that both phases are same.
    
    \item $SU(4)_2/\mathbb Z_2+2~\psi^{adj}=SO(6)_2+2~\psi^{adj}$
    
    We have $Q[SU(4)_2^{2adj}]=U(2)_{0,4}=U(1)_2$. Note that there is confined $\mathbb Z_2\in \mathbb Z_4$ 1-form symmetry\cite{Gaiotto:2014kfa} thus $Q[(SU(4)_2/\mathbb Z_2)^{2adj}]=U(1)_2\times SO(3)_0$ which has two vacua from the spontaneously broken $\mathbb Z_2$ magnetic symmetry \cite{Kapustin:2014gua}. Hence it is more safe to analyze from $SO(6)$ side rather than $Spin(6)$ theory. While the $Q[SO(6)^{2adj}]=Q[SO(2)^{2sym}]=U(1)_2$ from the naive duality chain, we commented on the footnote \ref{fn1} that $Q[SO(2)^{2sym}]$ case is exceptional since dual description doesn't preserve common faithful global symmetry. Remarkably, consistent picture can be obtained if $SO(2)+2 ~\psi^{sym}$ flows to $\mathbb C\mathbb P^1$ since then the double monopole deformation $\mathcal M^2$ which is required for the matching of global symmetry\cite{Cordova:2017kue} generically breaks $\mathbb C\mathbb P^1$ to north and south poles which gives $\mathbb Z_2$ vacua with minimal TQFT $U(1)_2$ saturating UV 1-form anomaly.

\end{itemize}
    
\subsection{Self Consistency under $SL(2,\mathbb Z)$ transformation} \label{subsec:sl(2,z)selfcons} 

For the unitary gauge group case, it is necessary to establish the consistency of the duality chain and $SL(2,\mathbb Z)$ transformation. We start by explicitly writing down the consistent coupling of $U(1)$ background gauge field for the $SU(N)_k+2~\psi^{adj}$ phase diagram in the figure \ref{fig:su(n)smallk}. Then we show that applying the $ST^2$ transformation to the phase diagram \ref{fig:su(n)smallk} of the original theory becomes similar to the first step of the duality chain in the figure \ref{fig:su(n)firstdualitychain} where the only difference is the shift of rank from $N$ to $N+k$. We comment about the subtlety of the overall background $U(1)_B$  counterterm, which resolves the naive contradiction when one applies the $SL(2,\mathbb Z)$ transformation\footnote{For the introduction to the $SL(2,\mathbb Z)$ transformations in 2+1 dimensional gauge theory, see \cite{Witten:2003ya}} to the general expression \eqref{eq:su(n)qphase} of the quantum phase of $SU(N)_k+2~\psi^{adj}$ when we go up the duality chain.

The Lagrangian density of the original theory together with left and right dual descriptions are as follows:
\beq \label{eq:su(n)originalbackgroundcoupling}
~&\mathcal L_{SU(N)_k[B]+2~\psi^{adj}}=\frac{k}{4\pi}Tr(ada-\frac{2i}{3}a^3)+\frac{1}{2\pi}ed(Tra +B)-\frac{1}{4\pi}BdB+\psi D_{a}\psi
\\~&\mathcal L_{U(N-k)_{k,N}[B]+2~{\hat \psi}^{adj}}=\frac{k}{4\pi}Tr(udu-\frac{2i}{3}u^3)+\frac{1}{4\pi}Tru d Tru - \frac{1}{2\pi}Tru dB+\hat \psi D_{u}\hat \psi
\\~&\mathcal L_{U(N+k)_{k,-N}[-B]+2~{\tilde \psi}^{adj}}=\frac{k}{4\pi}Tr(vdv-\frac{2i}{3}v^3)-\frac{1}{4\pi}Trv d Trv +\frac{1}{2\pi}Trv dB+\tilde \psi D_{v}\tilde \psi-\frac{2}{4\pi}BdB
\eeq

The coupling of $B$ and the choice of overall counterterm is consistent in the following sense. Trivially, it is consistent with level/rank duality\cite{Hsin:2016blu} at the two asymptotic phases. First, the left asymptotic phase $m_\psi \rightarrow
 -\infty$ where duality chain starts doesn't have any counterterm of $B$. Moreover, there is no direct coupling of $B$ with fermions, which means that this $U(1)_B$ gauge field is not coupled with a flavor symmetry but coupled with a topological U(1) current along the phase diagram.\footnote{$SU(N)$ Chern-Simons theory is conveniently represented by $U(N)$ gauge field constrained by $U(1)$ auxiliary field to become traceless. U(1) background field $B$ is coupled to this auxiliary field but not the dynamical field directly. See \cite{Hsin:2016blu} for the detail.} 

Now we see that two main features pointed out at the above paragraph make the $SL(2,\mathbb Z)$ transformation self-consistent to the phase diagram. If we apply the $ST^2$ to the \eqref{eq:su(n)originalbackgroundcoupling} simultaneously, we can directly see that the theories become $U(N)_{k,k+N}+2~\psi^{adj}$, $ST^2U(N-k)_{k,N}+2~{\hat \psi}^{adj}$, $SU(N+k)_k+2~{\tilde \psi}^{adj}$, which are nothing but the components of the first duality chain of the phase diagram generated by $SU(N+k)_k$ with 2 adjoint fermions similar to figure \ref{fig:su(n)firstdualitychain}. Thus we see that structure of the duality chain is preserved under $SL(2,\mathbb Z)$ transformations.

Furthermore, the overall choice of the counterterm proportional to $\frac{1}{4\pi}BdB$ resolves the possible potential contradiction. Consistentcy of the expression of $Q[SU(N)_k^{2adj}]$ in \eqref{eq:su(n)qphase} with duality chain requires that $Q[U(N+k)_{k,-N}^{2adj}]$ should give same answer since it is just one step above in the duality chain. Thus we could apply $SL(2,\mathbb Z)$ transformation to the $Q[SU(N+k)_k^{2adj}]$ case and to see whether it is consistent or not. Naively, the right transformation looks to be $ST^{-1}$ which gives a contradiction in the sense that phase doesn't match which should be automatically guaranteed by the algorithm of duality chain. This is resolved by the correct counterterms $-\frac{1}{4\pi}BdB$ in the original theory and $-\frac{2}{4\pi}BdB$ in the right dual theory written in \eqref{eq:su(n)originalbackgroundcoupling}. Thus the correct required operation is $T^{-2}S$ which directly cancels $ST^2$\footnote{$S^2=C$, but the action of $C$ to $B$ can be undone since we could redefine the sign of the abelian dynamical gauge field}. Thus we see that $Q[SU(N)_k^{2adj}]=T^{-2}SQ[SU(N+k)_k^{2adj}]=Q[U(N+k)_{k,-N}^{2adj}]$.

Finally, we can connect the duality chain of $SU(N)_k+2\psi^{adj}$ to the $U(N)_{k,k}+2\psi^{adj}$ in the section \ref{sec:more dualities} by the overall $S$ transformation to the \eqref{eq:su(n)originalbackgroundcoupling}. Then we get the simple duality chain $U(N-mk)_{k,k}+2\psi_m^{adj},~m\in \mathbb Z$ which shares common intermediate phase without need of any generalized level/rank dualities.

\subsection{Special and General Matching of Gravitational Counterterms} \label{subsec:gravitational}
    
    In general, consistency check with the various counterterms in 2+1 dimensions \cite{Closset:2012vp,Closset:2012vg} along the non-trivial closed paths in the phase diagrams are not able to achieve since one of the dual description in the phase diagram \ref{fig:su(n)smallk} to \ref{fig:sp(n)asymsmallk} is always strongly coupled. In this section, we focus on the gravitational counterterm $c$ which conventionally defined as the coefficient of the twice of the gravitational Chern-Simons term 2CS$_{grav}$ from the background Riemannian metric. In our case, there are two types of contribution to the gravitational counterterm. One is related to the parity anomaly \cite{Niemi:1983rq,Redlich:1983kn,Redlich:1983dv} of the fermions in the curved background, and the other is from the physical requirement of the same thermal conductivity along the level/rank duality \cite{Hsin:2016blu, Aharony:2016jvv}.

For the theory with rank-two/adjoint cases, it turns out that there are some special cases where one could see the non-trivial matching of gravitational counterterm along the phase diagram. One is the case of isomorphism between low-rank Lie groups as discussed in \ref{sec:specialcase} where there is a two inequivalent steps leads to the same intermediate phases which can be used to establish the consistency with various counter-terms. This could be done using the various results in the literatures, e.g. \cite{Hsin:2016blu,Aharony:2016jvv,Cordova:2017vab,Choi:2018tuh}.

Another one is the case of specific small Chern-Simons level of UV theory where the quantum phase of the duality chain matches with the semiclassical phase appears in the 2-dimensional phase diagram obtained by varying the masses of two flavors independently\footnote{2-dimensional phase diagram for the vector fermions was analyzed in detail in \cite{Argurio:2019tvw,Baumgartner:2019frr}}.

Surprisingly, for the case of the two bifundamental fermions which discussed in \ref{subsec:two-bifund}, we could show that gravitational counterterm is consistent along the whole strongly coupled region of duality chain. We now explicitly show the last two tests.

\medskip
\noindent
$\bullet$ {\it \text{Matching duality chain versus 2 dimensional diagram} }
\medskip

Here we use the simplest example of $SU(N)_1+\psi^{sym}+\psi^{asym}$ where we could find a non-trivial closed loop in the 2-dimensional phase diagram of $(m_{\psi^{sym}},m_{\psi^{asym}})$. Duality chain of original theory to the intermdiate phase consists of following :

\beq \label{eq:gravisymasym}
(ST^{-2})^{m-1}U(N-m)_{1,-N+m+1}+\psi^{sym}_{-m}+\psi^{asym}_{-m}, ~m=0\dots N-1
\eeq

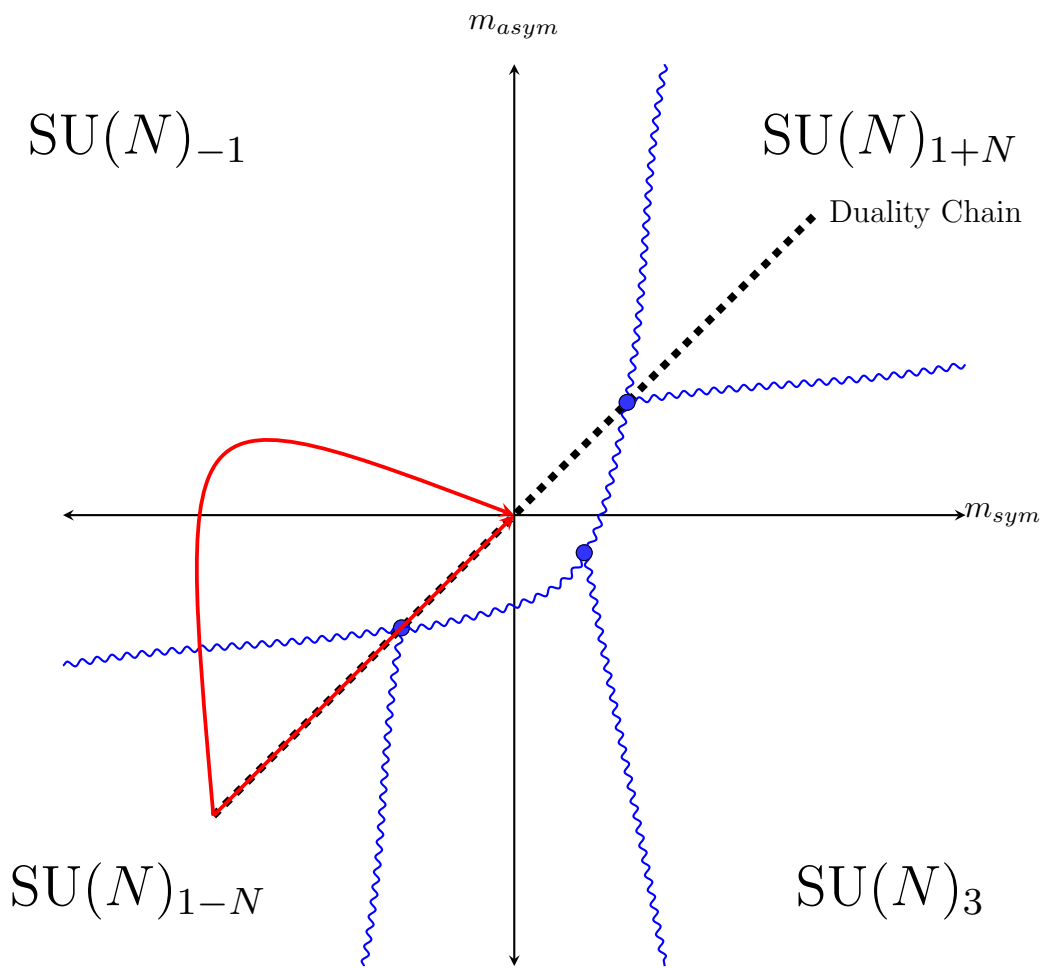
\begin{figure}[!h]
\centering
\begin{tikzpicture}

\draw[thick,<->,>=stealth] (0,0) -- (12,0);
\draw[thick,<->,>=stealth] (6,-6) -- (6,6);


\draw[blue,thick,decorate,decoration={snake,amplitude=.4mm,segment length=2mm}]  (8,6) to[out=-100,in=80] (7.5,1.5);
\draw[blue,thick,decorate,decoration={snake,amplitude=.4mm,segment length=2mm}]  (12,2) to[out=190,in=10] (7.5,1.5);

\draw[thick, dashed,line width=1 mm]  (7.5-5.5,1.5-5.5) -- (7.5+2.5,1.5+2.5) node[anchor=west] {Duality Chain};

\begin{scope}[rotate=180,shift={(-12,0)}]
\draw[blue,thick,decorate,decoration={snake,amplitude=.4mm,segment length=2mm}]  (8,6) to[out=-100,in=80] (7.5,1.5);
\draw[blue,thick,decorate,decoration={snake,amplitude=.4mm,segment length=2mm}]  (12,2) to[out=190,in=10] (7.5,1.5);
\filldraw[white!20!blue] (7.5,1.5) circle (3pt);\draw (7.5,1.5) circle (3pt);
\end{scope}

\draw[blue,thick,decorate,decoration={snake,amplitude=.4mm,segment length=2mm}] (7.5,1.5).. controls (7.0,-1.0) ..(4.5,-1.5);

\draw[blue,thick,decorate,decoration={snake,amplitude=.4mm,segment length=2mm}] (6.95,-0.5)to[out=-80,in=100] (8.0,-6.0);

\draw[red, ->,>=stealth,line width=0.5mm] (7.5-5.5,1.5-5.5) -- (6,0);
\draw[red, ->,>=stealth,line width=0.5mm] (7.5-5.5,1.5-5.5)  .. controls (1.5,1.7) .. (6,0);

\filldraw[white!20!blue](6.93,-0.5) circle (3pt);\draw(6.93,-0.5) circle (3pt);

\filldraw[white!20!blue] (7.5,1.5) circle (3pt);\draw (7.5,1.5) circle (3pt);

\node[scale=1.8] at (11,5) {$\mathrm{SU}(N)_{1+N}$};
\node[scale=1.8] at (1,-5) {$\mathrm{SU}(N)_{1-N}$};
\node[scale=1.8] at (1,5) {$\mathrm{SU}(N)_{-1}$};
\node[scale=1.8] at (11,-5) {$\mathrm{SU}(N)_{3}$};

\node[scale=1.2] at (12.5,0) {\footnotesize$m_{sym}$};
\node[scale=1.2] at (6,6.5) {\footnotesize$m_{asym}$};

\end{tikzpicture}
\caption{Generic shape of the two dimensional phase diagram for $SU(N)_1+\psi^{sym}+\psi^{asym}$. The diagonal black line represent the region where duality chain applies. Straight and curved red lines represent the two different paths connecting $SU(N)_{1-N}$ and the middle quantum phase by duality chain and 2d phase diagram analysis respectively. Blue wavy lines describe various phase transitions expected from a effective single adjoint theory at the asymptotic regions of the plane. The detailed shape of the critical lines at the bottom and right part is not important for the main discussion.}\label{fig:2dphase}
\end{figure}

Quantum phase from the duality chain is $(ST^{-2})^{N-2}U(1)_{-2}$ which is level rank dual to $SU(N)_{-1}$\cite{frohlich1994integral}, identical to the semiclassical phase with $(m_{\psi^{sym}},m_{\psi^{asym}})\rightarrow (-\infty, +\infty)$. Thus we expect the 2-dimensional phase diagram would look like figure \ref{fig:2dphase} where the middle quantum phase and 2nd quadrant is smoothly connected.

Now the consistency of the proposal requires that difference of the gravitational counterterm between the semiclassical phases near $(m_{\psi^{sym}},m_{\psi^{asym}})=(-\infty,-\infty)$ and the middle quantum phase region near $(m_{\psi^{sym}},m_{\psi^{asym}})=(0,0)$ shouldn't depend on the paths we choose to connect the two phases. Thus it following tests are necessary :

\beq
~& \Delta c_1[(m_{\psi^{sym}},m_{\psi^{asym}})~:~(-\infty,-\infty) ~\text{to}~(-\infty, +\infty) ]
\\&=\Delta c_2 ([\text{duality chain from } SU(N)_{1-N}\text{ to }SU(N)_{-1}])
\eeq

First, evaluation of $\Delta c_1$ is straightforward where we get $\Delta c_1=\frac{N(N-1)}{2}$. 

For the $\Delta c_2$, there are three different contributions to the gravitational counterterm jumps :

\begin{itemize}
\item N-1 times of generalized level rank duality at the each step of duality chain
\item N-1 times of mass deformation of $\psi_{-m}^{sym}$ and $\psi_{-m}^{asym}$, $m=1,...,N-1$ in \eqref{eq:gravisymasym}
\item Level/rank duality of $(ST^{-2})^{N-2}U(1)_{-2}$ to $SU(N)_{-1}$
\end{itemize}

Sum of the three contribution $\Delta c_2$ turns out to be the same as $\Delta c_1$\footnote{See \cite{Delmastro:2019vnj} for the calculation of the gravitational counterterm in the presence of abelian TQFT.} :

\beq
\Delta c_2=\sum_{m=1}^{N=1} (N-m+1)(N-m)-\sum_{m=1}^{N-1}(N-m)^2+0=\frac{N(N-1)}{2}
\eeq

Hence we see that duality chain is consistent with the prediction of 2d phase diagram based on the single flavor quantm phases. One can repeat the similar checks to the other kind of theories with the low Chern-Simons level where there are more semiclasscial critical lines at the asymptotic region compared to the generic case. In this case, the one of the semiclassical phases are matched with the quantum phase predicted by duality chain, and the non-trivial check could be done similar to the above analysis.

\medskip
\noindent
$\bullet$ {\it Matching duality chains of general two bifudnamental theories}
\medskip

Now we demonstrate that the duality chain with two bifundamental fermions admits more non-trivial test because of its distinguishing feature compared to the single gauge group with adjoint/rank-two fermions. This difference has already appeared in the case of single bifundamental versus single adjoint/rank-two QCD$_3$ as discussed in \cite{Aitken:2019shs}. Namely, the quantum phase of the bifundamental theory admits a scalar dual description with a bifundamental representation. This is a more natural duality than the fermionic dual description in the sense that it directly descends from the scalar dual description of the quantum phases of the fundamental fermions in QCD$_3$ \cite{Komargodski:2017keh} by gauging the global symmetry. We will use this scalar description to demonstrate the non-trivial consistency of the gravitational counterterms along the whole duality chains.

Let's describe the above idea more carefully with a specific example of the theory $SU(N_1)_{k_1}\times U(N_2)_{k_2}+2~\psi^{bifund}$ discussed in \ref{subsec:two-bifund}. Let's choose $\ceil{\frac{N_1}{k_2}}\leq \ceil{\frac{N_2}{k_1}}$ without loss of generality. The theory develops duality chain if $\vert k_1 \vert <N_2,~ \vert k_2 \vert  <N_1$ except $k_1=k_2=0$. Then the following sequence of dual theories appears when we go up or down of the chain, where we define $\mathcal C_m$ as the theory describing m-th chain similar to the figure \ref{fig:dualitychain} :

\beq \label{eq:fermionbifund}
~&\mathcal C_m ~:~ SU(N_1+mk_2)_{k_1}\times U(N_2+mk_1)_{k_2}+2~\psi_{m}^{bifund},~ m=-(\ceil{\frac{N_1}{k_2}}-1),...,+\infty\in \mathbb Z
\eeq

We emphasize again that the structure of the duality chain is directly coming from the self-consistency of the assumption about the appearance of the quantum phase as a certain condition between Chern-Simons levels and the rank of the gauge group. Thus dual theory with two bifundamentals develop the quantum phase generically thus requires multiple recursive steps to access the quantum phase. 

So far, the story is parallel to the adjoint/rank-two cases. But as we mentioned at the beginning, each step of the duality chain has dual scalar descriptions in the 2-node quiver case descend from the boson-fermion dualities in the fundamental matter case. Specifically, we can think of $SU(N_1+mk_2)_{k_1} \times U(N_2+mk_1)_{k_2} +2~\psi_m^{bifund}$ as coming from the following dualities \cite{Komargodski:2017keh}\footnote{One can interchange the role of first and second gauge group and perform the parllel analysis for the counter-terms.} :

\beq  \label{eq:bifundscalarungauged}
SU(N_1+mk_2)_{k_1}~+&~U(2N_2+2mk_1)~\psi_m^{fund},~~ m_{\psi}= m^{\pm crit}
\\
&\longleftrightarrow\qquad
\begin{cases}
U(N_2+(m+1)k_1)_{-N_1-mk_2}~+~(2N_2+2mk_1)\,\phi^{fund}_{m} &,~ m_{\phi}=0
\\ U(N_2+(m-1)k_1)_{N_1+mk_2}~+~(2N_2+2mk_1)\, \phi^{fund}_{m-1} &,~ m_{\phi}=0
\end{cases}
\eeq

Where we have quantum phase described by non-linear sigma model with target space $\mathcal M(2N_2+2mk_1,N_2+(m+1)k_1)$ with some WZ term. Also there is a $U(2N_2+2mk_1)$ global symmetry consistent with the phase whole phase diagram. Now we could gauge the $U(N_2+mk_1)\times U(N_2+mk_1)$ subgroup of $U(2N_1+2mk_1)$ to get the new dualities with 2-node quiver: 

\beq \label{eq:dualbfscalar}
~&SU(N_1+mk_2)_{k_1}\times  U(N_2+mk_1)_{k_2}~+~2~\psi_m^{bifund},~~ m_{\psi}= m^{\pm crit}
\\
&\longleftrightarrow\qquad
\begin{cases}
U(N_2+(m+1)k_1)_{-N_1-mk_2} \times U(N_2+mk_1)_{N_1+(m+1)k_2}~+~2~\phi_m^{bifund} &,~ m_{\phi_m}=0
\\ U(N_2+(m-1)k_1)_{N_1+mk_2}\times U(N_2+mk_1)_{-N_1-(m-1)k_2}~+~2~\phi_{m-1}^{bifund} &,~ m_{\phi_{m-1}}=0
\end{cases}
\eeq

We see that after gauging the original theory becomes the fermionic duality chain that has appeared in \ref{eq:fermionbifund}. The remarkable property of the above two mutually non-local dual descriptions with scalars is that they should share identical intermediate phase without any need of duality transformation. The reason is that two dual scalar descriptions in the original ungauged theory \eqref{eq:bifundscalarungauged} have identical non-linear sigma model without any duality transformation hence same background counterterms. Thus intermdiate quantum phase after gauging the global symmetry should have no difference in any background counterterms when described by either of dual descriptions with bifundamental scalars. 

Now we are ready to test whether the gravitational counter-terms along the duality chains are consistent using two bifundamental scalars description. We recall the notation in the figure \ref{fig:dualitychain} where the infrared phases of the m-th duality chain $\mathcal C_m$ in \eqref{eq:fermionbifund} are denoted as $\mathcal C_m^{+},\mathcal Q[\mathcal C_m],\mathcal C_m^{-}$ each represent the phases where fermions are large positive, masless , large negative respectively. If we choose the reference point $C_0^{-}$, then consistency of the duality chain requires that $\Delta c$ with respect to the intermediate quantum phase accessed from dual scalar description of the $\mathcal C_m$ in  \eqref{eq:dualbfscalar} should be independent of m, which we call it as $\Delta c [ \mathcal C_0^{-} \xrightarrow{duality ~chain} \mathcal C_m^{\pm } \xrightarrow {scalar ~dual} \mathcal Q[\mathcal C_m]=0 ]$. Surprisingly, it turns out that counterterms are consistent as follows\footnote{Note that the case for the orthogonal and symplectic groups could be analyzed similarly and pass the same consistency check.}:

\beq
~&\Delta c \left[ \mathcal C_0^{-} \xrightarrow{duality ~chain} \mathcal C_m^{\pm } \xrightarrow {scalar ~dual} \mathcal Q[\mathcal C_m]\right ]=N_1(N_2-k_1)=const ~!~
\eeq

\section{Discussion and Future Directions} \label{sec:comments}

In this paper, we propose a way to analyze the phase diagrams of QCD$_3$ with two fermions under any combination of adjoint/rank-two representations, or two bifundamental fermions using the duality chain. This was achived by realizing the possibility that dual description could be `strongly coupled' in the sense that dual description could also develop non-perturbative quantum phases.

Thus we hope that the concept of strong-strong duality and duality chain would lead to further understanding of strongly coupled dynamics in QCD$_3$. We list here several possible future directions:

\medskip
\noindent
$\bullet$ {\it IR phase for $k=0$}
\medskip

In the main text, we neglected the discussion of the phase diagram when the Chern-Simons level $k$ is zero. Existence of the quantum phase for the case of $SU(2)+2~\psi^{adj}$ was anticipated in \cite{Choi:2018tuh}, which was the gapless phase originated from the spontaneous symmetry breaking of $SO(2)$ flavor symmetry. Thus it is natrual to expect two adjoints theory with $k=0$ would similary has a identical gapless phase if we follow the logic of \cite{Choi:2018tuh}. Moreoever, it is also important to find a candidate IR phases for the $k=0$ in the case of any pair of rank-two/adjoints fermions or two bifundamental fermions treated in \ref{sec:more dualities}. We hope to return to this in the future.

\medskip
\noindent
$\bullet$ {\it Generalization to the higher $N_f$ or higher representations}
\medskip

The extension of duality chain to the case of general number of matters or higher representation is also an interesting direction. Evidence for the quantum phase in higher $N_f$ follows from the observation that $SU(2)+2 ~\psi^{sym}$ analyzed in the section \ref{sec:more dualities} is same as $SU(2)+4~\psi^{adj}$ with $SU(2)$ flavor symmetry preserving deformations. As in the $N_f=2$ adjoints case\cite{Anber:2018tcj,Bi:2018xvr,Cordova:2018acb,Wan:2018djl}, the intuition from 3+1 adjoint QCD with higher $n_f$ would be helpful, e.g. \cite{Poppitz:2019fnp}. The existence of the quantum phase for the matter with higher representations are also positive from the analysis of the main text. For instance, $SU(4)+\ydiagram{2}$ is equivalent to $Spin(6)+\ydiagram{1,1,1}$, or $Spin(5)+\ydiagram{2}$ is equivalent to $Sp(2)+\ydiagram{2,2}$. It is interesting to find out to what extent the quantum phase or duality chain exist.	

\medskip
\noindent
$\bullet$ {\it Construction of two dimensional phase diagrams}
\medskip

Focusing on the theories discussed in this paper where we always have two flavors of fermion, it is also natural to ask how does the two dimensional diagram with flavor symmetry breaking masses would look like. Similar work for the fundamental matter was done in \cite{Argurio:2019tvw,Baumgartner:2019frr}. One notable difference is the case when the adjoint fermion is included, since then we have to answer how does the critical lines of Goldstino look like coming from the spontaneous SUSY breaking \cite{Witten:1999ds}. Careful construction of two-dimensional phase diagram might give a non-trivial test of our proposal.

\medskip
\noindent
$\bullet$ {\it Orbifold Equivalence}
\medskip

Regarding the more nontrivial evidences coming from two bifundamental theories in \ref{subsec:gravitational} (by this we mean the matching of gravitational counterterm.), it is also natural to ask whether we could connect the two bifundamental theories to the theory of two adjoint fermions. The answer was positive for the case of single bifundamental and single adjoint fermions in \cite{Aitken:2019shs} using the orbifold projection. Similar technique gives a positive answer between two bifundamentals and two adjoints by using similar $\mathbb Z_2$ automorphisms along the dualities to single flavor case. Thus it is natural to conjecture some kinds of 2+1 dimensional version of planar equivalence should hold between two theories. It would be nice to see more evidence of this connection.

\section*{Acknowledgments}

We would like to thank D.~Delmastro, J.~Gomis, D.~Radi\v cevi\'c, M.~Roček, S.~Seifnashri, A.~Sharon, M.~Yu for useful discussions. We would also like to especially thank Z.~Komargodski for the invaluable support, discussions and the careful reading of this manuscript. CC is supported in part by the Simons Foundation grant 488657 (Simons Collaboration on the Non-Perturbative Bootstrap). Any opinions, findings, and conclusions or recommendations expressed in this material are those of the authors and do not necessarily reflect the views of the funding agencies.

\begin{appendix}
\section{Generalized Level-Rank Duality} \label{appsec:genlevelrank}

In this appendix, we derive the level-rank dual of $U(N)_{K,K\pm pN}$ using $SL(2,\mathbb Z)$ transformations on the standard $SU(N)_K \leftrightarrow U(K)_{-N,-N}$ level-rank duality. We start from showing that $U(N)_{K,K\pm N}\leftrightarrow U(K)_{-N,\mp k -N}$ is directly obtained by $SU(N)_K \leftrightarrow U(K)_{-N}$ with simple $SL(2,\mathbb Z)$ operation. 

First we review the elements of $SL(2,\mathbb Z)$ transformation for the 3d QFT with $U(1)$ global symmetry coupled to the background field $B$, following the notation of \cite{Hsin:2016blu}. If we denote the Lagrangian density of such theory as $\mathcal L[B]$, $S$ and $T$ trasnformation is defined as follows \cite{Witten:2003ya} :

\beq
~&S ~:~\mathcal L[B]\rightarrow \mathcal L[b]-\frac{1}{2\pi}bdB
\\&T~:~ \mathcal L[B]\rightarrow \mathcal L[B]+\frac{1}{4\pi}BdB
\eeq

Where the $b$ field in the $S$ operation is a dynamical field from the gauging of the original $U(1)$ background field. The $S$ and $T$ operations generate the $SL(2,\mathbb Z)$ operation on the theory space of 3d QFT with $U(1)$ global symmetry, satisfying following identities : $S^2=(ST)^3=\mathcal C$.

Now we discuss how this $SL(2,\mathbb Z)$ operation derives the generalized level/rank dualtiy for the unitary gauge group. We start from the seed level/rank duality $SU(N)_K \leftrightarrow U(K)_{-N,-N}$ 3d Chern-Simons TQFT with the following form of consistent mapping between the $U(1)$ baryon/monopole background gauge field $B$ on the each side \cite{Hsin:2016blu} :

\beq \label{app:SUU}
&\mathcal L_{SU(N)_K[B]}=\frac{K}{4\pi}Tr_N[bdb-\ti b^3]+\tp cd(Trb +B) 
\\&\longleftrightarrow \mathcal L_{U(K)_{-N,-N}[B]}=\frac{-N}{4\pi}Tr_K[udu-\ti u^3]+\tp (Trb) dB
\eeq

Now $U(N)_{K,K\pm N} \leftrightarrow U(K)_{-N,\mp K -N}$ is derived once we apply the operation $ST^{\pm 1}$ on the both side of \eqref{app:SUU} as follows :

\beq
& \mathcal L_{ST^{\pm 1}SU(N)_K[B]} =\frac{K}{4\pi}Tr_N[bdb-\ti b^3]+\tp cd(Trb +e)\pm \fp ede-\tp edB
\\&\longleftrightarrow  \mathcal L_{ST^{\pm 1}U(K)_{-N,-N}[B]}=\frac{-N}{4\pi}Tr_K[udu-\ti u^3]+\tp (Tru) de \pm \fp ede-\tp edB
\eeq

After integrating out $c$ and $e$ we obtain $\mathcal L_{U(N)_{K,K\pm N}[B]} \leftrightarrow  \mathcal L_{U(K)_{-N,\mp K-N}[\pm B]}+\mathcal L_{U(1)_{\pm 1}[-B]}$, thus we get the desired duality $U(N)_{K,K\pm N} \leftrightarrow U(K)_{-N,\mp K -N}$. Note that this simplification only happens to this case. More general level-rank duality is obtained once we apply $ST^p$ to the both side of \eqref{app:SUU} :

\beq
& \mathcal L_{ST^{p}SU(N)_K[B]}=\frac{K}{4\pi}Tr_N[bdb-\ti b^3]+\tp cd(Trb +e)+\frac{p}{4\pi} ede-\frac{1}{2\pi} edB \\& \longleftrightarrow \mathcal L_{ST^{p}U(K)_{-N,-N}[B]}=\frac{-N}{4\pi}Tr_K[udu-\ti u^3]+\tp (Tru) de + \frac{p}{4\pi} ede-\frac{1}{2\pi} edB
\eeq

After integrating out the auxiliary fields, we obtain following generalization of level-rank duality :

\beq \label{app:genlevelrank}
U(N)_{K,K+pN}[B]\qquad \longleftrightarrow \qquad ST^pU(K)_{-N,-N}[B]
\eeq

It is useful to write down the the generalized level-rank duality obtained similarly from time-reversal of \eqref{app:SUU}:

\beq \label{app:trevgenlevelrank}
U(N)_{-K,-K+pN}[B]\qquad \longleftrightarrow \qquad ST^pU(K)_{N,N}[-B]=ST^pU(K)_{N,N}[B]
\eeq

Where we used the fact that redefining the sign of abelian dynamical gauge field effectively makes $B$ to $-B$.

\end{appendix}

\printbibliography

\end{document}